\RequirePackage{fix-cm}
\documentclass[twocolumn,epjc3]{svjour3}  
\smartqed  
\RequirePackage{color,graphicx,url}

\journalname{Eur. Phys. J. A}

\usepackage{amsfonts,amsmath,amssymb,bm,xspace}
\graphicspath{{./figures/}}

\usepackage{hyperref}                

\usepackage{academicons} 
\newcommand{\orcid}[1]{\href{https://orcid.org/#1}{\textcolor[HTML]{A6CE39}{\aiOrcid}}}
\usepackage{xcolor}
\usepackage{hyperref}

\newcommand{\sat}{\mathrm{sat}}
\newcommand{\sym}{\mathrm{sym}}

\newcommand{\fmiq}{\, \text{fm}^{-3}}

\newcommand{\MeV}{\, \text{MeV}}

\renewcommand{\L}{\mathcal{L}}
\newcommand{\psib}{\bar{\psi}}

\newcommand{\QCD}{\ensuremath{{\rm QCD}}}

\usepackage{graphicx}

\hypersetup{%
    pdfsubject=Paper,
    pdfkeywords={nuclear physics} {MBPT} {chiral EFT} {asymmetric nuclear matter}
    unicode=true,
    breaklinks=true,
     colorlinks   = true,
     linkcolor = blue,
     citecolor = blue,
     menucolor = blue,
     urlcolor = blue
}

\begin{document}

\title{Comparison of different relativistic models applied to dense nuclear matter}


\author{Rahul Somasundaram\thanksref{addr1}\href{https://orcid.org/0000-0003-0427-3893}{\includegraphics[scale=0.5]{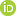}} 
        \and
        J\'er\^ome Margueron\thanksref{addr1}\href{https://orcid.org/0000-0001-8743-3092}{\includegraphics[scale=0.5]{figures/orcid.png}}
        \and
        Guy Chanfray\thanksref{addr1}\href{https://orcid.org/0000-0002-3593-9507}{\includegraphics[scale=0.5]{figures/orcid.png}}  
        \and
        Hubert Hansen\thanksref{addr1}\href{https://orcid.org/0000-0001-8879-3612}{\includegraphics[scale=0.5]{figures/orcid.png}} 
}

\institute{Univ Lyon, Univ Claude Bernard Lyon 1, CNRS/IN2P3, IP2I Lyon, UMR 5822, F-69622, Villeurbanne, France \label{addr1}}


\date{Draft version:\today, Received: date / Accepted: date}

\maketitle

\abstract{We explore three different classes of relativistic approaches applied to the description of dense nuclear matter: a Walecka-type relativistic mean field model (RMF), an extension including an effective chiral potential (RMF-C) and a further extension with a chiral potential and confinement effects (RMF-CC).
The parameters of the latter are controlled by fundamental properties such as the chiral potential, Lattice-QCD predictions, the quark sub-structure, as well as empirical properties at nuclear matter saturation. 
While these models are calibrated to the same properties at saturation density, they differ in their predictions as the density increases. We take care of parameter uncertainties and propagate them to our predictions for symmetric nuclear matter by employing Bayesian statistics. We show that RMF and RMF-C share common features as the density increases, while RMF-CC behaves differently. For instance, the scalar field at 6$n_\sat$ reaches $\sim 20$~MeV for RMF-CC while it is larger than $\sim 70$~MeV for RMF and RMF-C.
Interestingly, we also show that, by fixing the $\rho$ coupling constant from the quark structure of the nucleon, these three models reproduce only half of the empirical symmetry energy.}
 
\PACS{
{12.39.Fe}{Chiral Lagrangians}
\and
{21.65.+f}{Nuclear matter}
\and
{26.60.-c}{Nuclear matter aspects of neutron stars}
} 

\section{Introduction}
\label{sec:intro}

The understanding of the properties of dense nuclear matter remains a challenge since, on the theoretical side Quantum Chromo-Dynamics (QCD) cannot be solved directly and, on the experimental side very few data exist, see Ref.~\cite{CBM2017} for a recent review. Effective nuclear modeling may be employed to tackle the problem and efforts had been made to connect those descriptions to the fundamental theory QCD, in particular its chiral properties~\cite{Weinberg}. This effective approach is well suited to address low-energy systems, but as the energy, or equivalently the density, increases it faces a natural break-down. In cold neutron matter, the break-down is expected to occur between $n_\sat$ and $2n_\sat$, where $n_\sat$ is the nuclear saturation density ($n_\sat\approx 0.155$~fm$^{-3}$), see for instance Ref.~\cite{Tews:2018kmu} and references therein. The exploration of the densest phase of nuclear matter therefore requires extrapolations such as the one proposed in Ref.~\cite{Hebeler2013}.

In this study, we adopt a different viewpoint and investigate a relativistic modeling of nuclear matter along the lines originally proposed in Refs.~\cite{Chanfray2001,Chanfray2005} where spontaneous chiral symmetry breaking and confinement effects are incorporated (RMF-CC model). We also compare this model to other effective approaches such as the Walecka-type RMF model~\cite{SerotWalecka1986,Walecka1997} and the RMF-C ones, inspired from Refs.~\cite{Wetterich,Drews2013,Fraga2018,Schmitt2020} (note that in the present work we introduce the names RMF-CC and RMF-C for convenience). We focus our analysis on relativistic frameworks since such approaches aim at describing the dense core of neutron stars (NSs). In particular, recent radio observations~\cite{Demorest:2010,Antoniadis:2013pzd,Cromartie:2019,Fonseca:2021wxt} as well as X-ray observations from NICER~\cite{Miller:2021qha,Riley:2021pdl} of NSs with masses around two solar masses have indicated that the sound speed in the cores of NSs is expected to be larger than 10\% of the velocity of light~\cite{Tews:2018kmu,Bedaque:2014sqa}. Therefore the study of the densest phases in the core of neutron stars requires the development of a relativistic description of nuclear matter.

While for densities well above the saturation density, the question of possible phase transitions and natural degrees of freedom is 
important, see Ref.~\cite{Somasundaram:2021clp} for instance, in the present paper we restrict ourself to matter made of nucleons only. The model dealing with the nucleonic substructure (RMF-CC),
however, already contains the seeds for the emergence of more fundamental degrees of freedom and could be extended -- in the future -- for a description of phase transitions in dense matter.
Bridging the fundamental aspects of QCD with nucleonic degrees of freedom has indeed became a fundamental question in the recent development of nuclear physics. Typically around saturation density, one could consider nucleons as the basic degrees of freedom that exist in nuclei since energies up to few tens of MeV per nucleon are involved~\cite{Wong:1998ex}. By the advent of QCD, it has however established that nucleons are not fundamental, but they are instead composed of colored quarks which interact through the exchange of colored gluons at a resolution scale of the few hundreds of MeV per nucleon~\cite{Yndurain2006}. The typical low energy nuclear scale therefore remains too low to excite the nucleon substructure. This scale ordering has been employed to develop an effective theory of QCD at low energy~\cite{Weinberg}. 
Further, the quark sub-substructure could contributes to the polarization of the nucleon, which in turn might play an important role in the properties of dense matter~\cite{Guichon1988}.

Before returning to this point, a brief reminder of the development of our understanding of the low energy nuclear interaction is necessary. 
In 1935, the Yukawa meson-exchange model~\cite{Yukawa1935} has
shaped the global understanding of nuclear physics in terms of nucleons and mesons, producing the first good qualitative results by fixing the coupling constant and associating a particle exchange to the strong interaction. 
In the early 1960’s, the discovery of heavy mesons helped in the modeling of better one-boson-exchange potentials (OBEP) containing the exchange of well identified vector mesons namely the omega ($\omega$) and rho ($\rho$) mesons. There were however still some problems, e.g., the scalar sigma boson-exchange, now named $f_0(600)$, for which the experimental evidence was polemic as well as its link to the broad 2$\pi$ scalar resonance. Nevertheless, high-precision potentials based on the meson-exchange picture with the inclusion of a scalar meson were constructed and successful, see for instance Ref.~\cite{Erkelenz1971}.

Since then, some phenomenological approaches such as the Walecka model~\cite{SerotWalecka1986,Walecka1997}, aimed at describing the binding energy in finite nuclei, as well as low-energy excitations, have anchored their modeling into the meson-exchange picture and have suggested that relativistic descriptions could explain both the nuclear saturation and the spin-orbit coupling.
It is nowadays often called the relativistic mean field (RMF) model, 
see Ref.~\cite{Lalazissis1996} for instance. 
The link between such Lagangians and the bare nucleon-nucleon interaction could be performed through the Dirac-Brueckner-Hartree-Fock (DBHF) approach~\cite{Brockmann1984,TerHaar1986} which produces a mean-field that guides the parametrization of the RMF model, introducing density-dependent coupling constants~\cite{Typel1999,vanDalen2011}. Some recent RMF models accurately reproducing nuclei properties have included this link to DBHF potentials~\cite{Lalazissis2005,Long2007}. 
However, the question of the very nature of the background mesonic fields is still to be elucidated or, said differently, it is highly desirable to clarify their relationship with the low-energy realization of the symmetries of QCD. 

Chiral symmetry together with color confinement are the most prominent low-energy features of QCD. In the limit of vanishing quark masses, the QCD lagrangian has essentially no dimensional parameter. This scale invariance is however broken by quantum fluctuations \cite{Yndurain2006} leading to the formation of the characteristic QCD momentum scale $\Lambda_{\QCD}\approx 200$~MeV. Much below $\Lambda_{\QCD}$, the coupling constant of the theory becomes very large, a feature which is supposed to generate color confinement and consequently render QCD a non-perturbative theory in the energy range of nuclear physics, often referred to as low-energy~\cite{Ripka1997}. Furthermore, the chiral symmetery between left and right handed quarks is spontaneously broken by the ground state of QCD leading to the formation of 'Goldstone bosons' (that are identified with pions) as well as their chiral partner, a scalar-isoscalar field.
To bridge the gap between relativistic theories of the Walecka type and approaches based on chiral symmetry, one has to map the \emph{nuclear physics} sigma meson of the Walecka model at the origin of the nuclear binding with a chiral quantity. In this work, we will review the argument suggested by Chanfray et al.~\cite{Chanfray2001}, complemented with the nucleon response~\cite{Chanfray2005}, to perform this mapping. 
We will also comment on and provide comparisons with other approaches. 


The paper is organized as follows. In Sec.~\ref{sec:lagrangian} the relativistic models are described in details, including the discussion of the link between the parameters and the fitted data as well as their uncertainties. For simplicity, we consider the Hartree approximation, also called the classical field case, and the properties of these models are explored in symmetric matter (SM), even though the question of the prediction of the symmetry energy is also addressed at the end of our study. Note that the parameter adjustment is done in such a way that all models are consistent at saturation density and extrapolation to high density is performed with Bayesian statistics. The predictions at high density therefore incorporate uncertainties from the model parameters. In Sec.~\ref{sec:comaparison} we begin the comparison of the relativistic models. In Sec.~\ref{sec:energy} we show that the predictions of the models do not agree at high density by studying the energy per particle and the self-energies. Secs.~\ref{sec:potential} and~\ref{sec:eom} focus on the interpretation of the scalar potential of the different models. Sec.~\ref{sec:Landau} is devoted to a discussion of the Landau parameter $F_0$.
Finally, in Sec.~\ref{sec:symmetry_energy} we discuss the predictions for the symmetry energy, and describe limitations of the Hartree approximation and possible ways to cure those limitations.

\section{The relativistic nuclear models}
\label{sec:lagrangian}

Considering only the lightest $u$ and $d$ quarks and the flavor number $N_f=2$, the chiral fields associated to the
fluctuations of the quark condensate $\langle \bar q q\rangle$ resulting from chiral symmetry breaking are usually parametrized in term of a $\rm{SU}(2)$ matrix $M$ as:
\begin{equation}
M=\sigma + i\vec{\tau}\cdot\vec{\phi}\equiv S\, U
\label{REPRES}
\end{equation}
with $S = s + f_\pi$ and $U=e^{i\,{\vec{\tau}\cdot\vec{\pi}}/{f_\pi}}$.
The scalar field $\sigma$ ($S$) and pseudoscalar fields $\vec{\phi}$ ($\vec{\pi}$) written in cartesian (polar) coordinates  appear as the dynamical degrees of freedom. As stated in Sec.~\ref{sec:intro}, it is necessary to clarify the connection between the nuclear physics sigma meson of the Walecka model (let us call it $\sigma_W$ from now on) at the origin of the nuclear binding with a chiral field~\eqref{REPRES}. For instance, one may be tempted to identify $\sigma_W$ with the scalar field $\sigma$ in cartesian coordinates. It is however forbidden by chiral constraints and this point has been first addressed by Birse~\cite{Birse94}: it would lead to the presence of terms of order $m_\pi$ in the NN interaction which is not allowed. 

In this study, we follow Ref.~\cite{Chanfray2001} and identify $\sigma_W$ with the chiral invariant $s$ ($=S-f_\pi$) field associated with the radial fluctuation of the chiral condensate $S$ around the \emph{chiral radius} $f_\pi$, in polar coordinates. It formally consists of promoting the chiral invariant scalar field $s$ and the pion field $\vec{\pi}$ appearing in the matrix $M$~\eqref{REPRES}
to effective degrees of freedom. This was originally formulated in the framework of the linear sigma model~\cite{Chanfray2001} but an explicit construction using a bosonization technique of the chiral effective potential can be done within the NJL model~\cite{Chanfray2011} where the linear sigma model potential is recovered through a second order expansion in  $S^2-f^2_\pi$ of the constituent quark Dirac sea energy.
This proposal, which gives a plausible answer to the long standing problem of the chiral status of Walecka theories, has also the merit of respecting all the desired chiral constraints~\cite{Birse94}. In particular the correspondence $s\equiv \sigma_W$ generates a coupling of the scalar field to the derivatives of the pion field, as expected in the physical world. Hence the radial mode decouples from low-energy pions whose dynamics is governed by chiral perturbation theory. A detailed discussion of this sometimes subtle topic is given in \cite{Chanfray2001,Chanfray2006}. 

Once the effective degrees of freedom are identified,
the relativistic Lagrangian can generically be written as the sum of a kinetic fermionic term,
\begin{equation}
\L_\psi = \psib \left( i \gamma^{\mu} -M_N \right)\partial_{\mu} \psi \, , \nonumber \\
\end{equation}
where the field $\psi$ represents the nucleon spinor, and of meson-nucleon interaction terms, 
\begin{equation}
\L_{m} = \L_{s} + \L_{\omega} + \L_{\rho} + \L_{\delta} + \L_{\pi} \, ,
\end{equation}
collecting all mesonic contributions considered in a given model. Using notation of Ref.~\cite{Massot2008} these can be enumerated as,
\begin{eqnarray}
\label{eq:L_meson}
\L_s &=& \big(M_N - M_N(s)\big) \bar{\psi} \psi - V(s) + \frac{1}{2} \partial^\mu s \partial_\mu s\, , \nonumber \\
\L_\omega &=& -g_\omega \omega_\mu \bar{\psi} \gamma^\mu \psi + \frac{1}{2} m_\omega^2 \omega^\mu \omega_\mu - \frac{1}{4} F^{\mu \nu} F_{\mu \nu}\, , \nonumber \\
\L_\rho &=& -g_\rho \rho_{a \mu} \bar{\psi} \gamma^\mu \tau_a \psi + g_\rho \frac{\kappa_\rho}{2 M_N} \partial_\nu \rho_{a \mu} \bar{\psi} \sigma^{\mu \nu} \tau_a \psi \nonumber \\
&&+ \frac{1}{2} m_\rho^2 \rho_{a \mu} \rho_a^\mu - \frac{1}{4} G_a^{\mu \nu} G_{a \mu \nu}\, , \\
\L_\delta &=& -g_\delta \delta_a \bar{\psi} \tau_a \psi - \frac{1}{2} m_\delta \delta_a \delta_a + \frac{1}{2} \partial^\mu \delta_a \partial_\mu \delta_a\, , \nonumber \\
\L_\pi &=& \frac{g_A}{2 f_\pi} \partial_\mu \varphi_{\pi a} \bar{\psi} \gamma^\mu \gamma^5 \tau_a \psi - \frac{1}{2} m_\pi^2 \varphi_{\pi a} \varphi_{\pi a} \nonumber \\
&&+ \frac{1}{2} \partial^\mu \varphi_{\pi a} \partial_\mu \varphi_{\pi a}\, , \nonumber
\end{eqnarray}
where the symbols have their usual meaning. 
In Eq.~\eqref{eq:L_meson}, two quantities are of particular interest to us, the scalar potential $V(s)$ and the $s$-field dependent nucleon mass $M_N(s)$. Different expressions for these quantities have been employed in the past. 
For instance, the RMF-CC approach, that we will detailed hereafter, employs the chiral potential breaking the chiral symmetry in the vacuum, while in RMF-C, the chiral potential is considered as an effective potential, which is adjusted to reproduce saturation properties.
Finally, the scalar potential $V(\sigma_W=s)$ in RMF is possibly non-linear and has been introduced in a pragmatic way to better reproduce the incompressibility modulus and the effective mass.

In this paper we will restrict our attention to only symmetric matter (SM), although the symmetry energy is explored at the end of our study. The energy density at the Hartree level can be computed from the Lagrangian of Eq.~\eqref{eq:L_meson} in the usual way \cite{Chanfray2005}. It is expressed as
\begin{eqnarray}
 \varepsilon &=&\int\frac{4 d^3 k}{(2\pi)^3}\left( \sqrt{k^2+M_N^2(s)} \,+\,g_\omega\,\omega_0\right) \,\Theta(k_F-k)\nonumber\\
 && +\,V(s)\,-\,\frac{1}{2}\,m^2_\omega\,\omega_0^2, 
\end{eqnarray}
where the scalar and vector fields are obtained from the equations of motion given in~\ref{sec:a1}.
Note that in relativistic approaches, two densities are defined, the vector density (or baryonic density) $\rho\equiv\langle \bar{\psi} \gamma^0  \psi\rangle$ and the scalar density
$\rho_S\equiv\langle \bar{\psi} \psi\rangle$.

Note that Eq.~\eqref{eq:L_meson} contains three isovector coupling constants, $g_\rho$, $g_\delta$ and $g_A$. Since most of the paper is dedicated to SM, these parameters play no role (at the Hartree level) and are included in Eq.~\eqref{eq:L_meson} only for the sake of completeness. In Sec.~\ref{sec:symmetry_energy} of the paper, where the symmetry energy is discussed, we will focus only on the role of the $\rho$ meson and set $g_\delta = 0$ since the effects of the $\delta$ are expected to be small~\cite{Massot2008}, and $g_A$ does not contribute at the Hartree approximation.
Furthermore, the masses of all the mesons, except the $\sigma$, are taken from hadron phenomenology. It should however be noted that, in the Hartree approximation, only the ratio of the coupling constant to the mass (for all mesons) determines our results and not the masses themselves.

In the next subsections, we present the following models: RMF-CC, RMF-C and RMF. 
Since RMF-CC is the model anchored into microscopic predictions from QCD, in part, we present it first. The two other models (RMF-C and RMF) will naturally be better understood if presented after, especially since they are calibrated to reproduce the same empirical parameters as the ones predicted by RMF-CC.

\subsection{Relativistic Mean Field including Chiral potential and Confinement effects (RMF-CC) }

\begin{table}[tb]
    \begin{center}
    \setlength{\tabcolsep}{15pt}
    \renewcommand{\arraystretch}{1.5}
    \caption{
    Nuclear Empirical Parameters (NEP) ($E_{\sat}$ and $n_{\sat}$) as given from Ref.~\cite{Margueron2018} and Lattice parameters ($a_2$ and $a_4$) extracted from Ref.~\cite{LWY2003}, used in the fits. For the NEP the mean and standard deviations correspond to a Gaussian distribution, while for the Lattice parameters the standard deviation refers to the width of a uniform distribution.}
    \label{tab:param}
    \begin{tabular}{ccc}
    \hline\hline
    Parameter &	Mean &	 Standard deviation\\
    \hline
    $E_{\sat}$ (MeV) & $-15.8$ & $0.3$ \\
    $n_{\sat}$ ($\fmiq$) & $0.155$ & $0.005$ \\
    $a_2$ (GeV$^{-1}$) & $1.533$ & $0.136$ \\
    $a_4$ (GeV$^{-3}$) & $-0.509$ & $0.054$ \\
    \hline\hline
    \end{tabular}
\end{center}
\end{table}

We first discuss the relativistic model that incorporates both chiral symmetry and some effects of confinement, namely the nucleon polarisability originating from its substructure treated in the linear response approximation.
The former is enunciated via the chiral effective potential, which is a Mexican hat potential for the scalar field,
\begin{equation}
\label{eq:V(s)}
V(s) = \frac{m_\sigma^2 - m_\pi^2}{8 f^2_\pi}\left(\sigma^2 +\phi^2-v^2 \right)^2\,-\,f_\pi m^2_\pi\sigma\, ,
\end{equation}
with
\begin{equation}
v^2= f^2_\pi\,\frac{m_\sigma^2 -3 m_\pi^2}{m_\sigma^2 - m_\pi^2}\, .
\end{equation}
The chiral effective potential~\eqref{eq:V(s)} corresponds to the original linear sigma model using the \emph{cartesian} coordinates, namely the chiral partners $\sigma$ and $\phi$.

\begin{figure*}
    \centering
    \includegraphics[width=0.8\textwidth,height=0.6\textwidth]{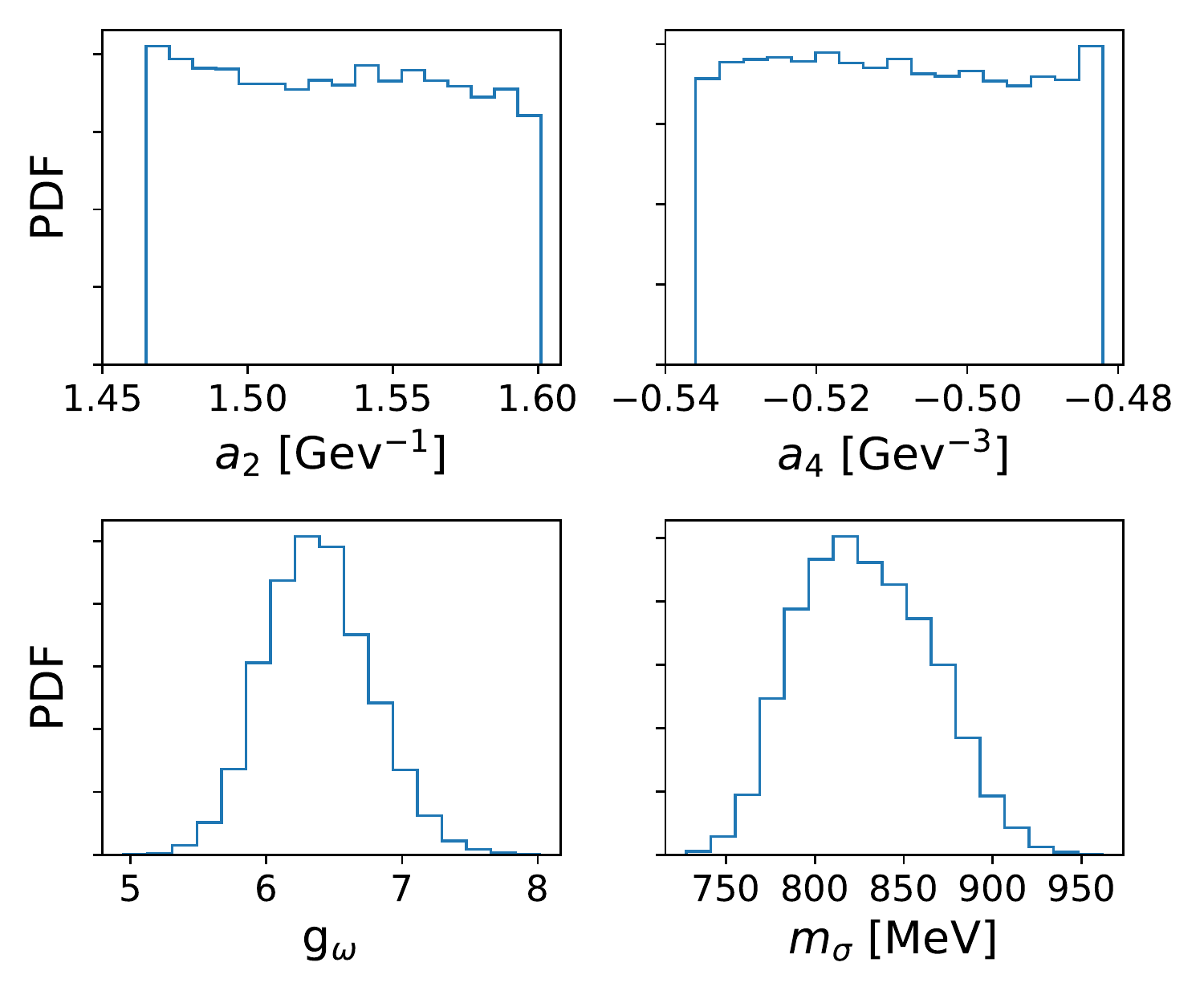}
    \caption{Probability Distribution Function (PDF) for the parameters of the RMF-CC model, adjusted to reproduce the NEPs $E_\sat$ and $n_\sat$ as well as the Lattice parameters $a_2$ and $a_4$, see Table~\ref{tab:param} for more details.}
    \label{fig:LSM}
\end{figure*}

As explained in Sec.~\ref{sec:lagrangian}, we employ the polar coordinates and, additionally, keep only the leading order mass term for the pion. The chiral potential can then be expressed as
\begin{eqnarray}
V(s) &=& \frac{m_\pi^2}{2}\left(S^2-f^2_\pi \right)\,+\,\frac{m_\sigma^2 - m_\pi^2}{8 f^2_\pi}\left(S^2-f^2_\pi \right)^2 \, \nonumber \\ 
&& \hspace{1cm} +\frac{1}{2} m_\pi^2\pi^2+...
\label{eq:V(s)_exp}
\end{eqnarray}

In Eq.~\eqref{eq:V(s)_exp} the higher order terms generate pion-pion interactions which disappear in the chiral limit.  

We finally get the following expression that will be used for this model:  
\begin{equation}
\label{eq:V(s)_s}
V(s)= \frac{m_\sigma^2}{2} s^2 + \frac{m_\sigma^2 - m_\pi^2}{2 f_\pi} \left( s^3 + \frac{s^4}{4 f_\pi} \right)\, ,
\end{equation}
where we only keep the radial fluctuation field $s$, the field identified with $\sigma_W$ the "nuclear physics" sigma meson.

We now come to a very important point. The potential~\eqref{eq:V(s)_s} does not allow for SM to saturate because of the attractive contribution of the $s^3$ term in Eq.~\eqref{eq:V(s)_s}, i.e., the tadpole diagram~\cite{Boguta83,KM74,BT01,C03}.
This problem can be circumvented by introducing the nucleon response to the scalar field at finite density, which is the central ingredient of the quark-meson coupling model (QMC), introduced in the seminal work of P. Guichon \cite{Guichon1988} and successfully applied to finite nuclei with an explicit connection to the Skyrme force \cite{Guichon2004}. The physical motivation to introduce this nucleonic response is the observation that nucleons experience huge fields at finite density, e.g. the scalar field is of the order of a few hundred of MeV at saturation density. Nucleons, being in reality composite objects, will react against the nuclear environment (i.e., the background nuclear scalar fields) through a (self-consistent) modification of the quarks wave functions. 
This effect may generate a three body force which brings the desired repulsion 
if confinement dominates spontaneous chiral symmetry breaking, as discussed in Ref. \cite{Chanfray2011} within particular models.
This is the key ingredient of the saturation mechanism of the RMF-CC model. The attractive chiral $s^3$ tadpole diagram responsible for the instability of the ground state at finite density is counterbalanced by the nucleon response driven by the susceptibility $\kappa_{NS}$ modifying the nucleon in-medium mass as:
\begin{equation}
\label{eq:M_N(s)}
M_N(s) = M_N + g_s s + \frac{1}{2} \kappa_{NS} \left(s^2 + \frac{s^3}{3 f_\pi} \right) \, .
\end{equation}
In Eq.~\eqref{eq:M_N(s)}, the quadratic term leads to a non-zero nucleon susceptibility, and we have added a cubic term, which is fixed such that the susceptibility
\begin{equation}
\kappa_{NS}= \frac{d^2 M_N(s)}{ds^2}
\end{equation}
vanishes at full chiral restoration~\cite{Chanfray2005}. In the following, it will be convenient to introduce the dimensionless quantity 
\begin{eqnarray}
C = \frac{f_\pi^2}{2M_N} \kappa_{NS}\, .
\end{eqnarray}

\begin{figure*}
    \centering
    \includegraphics[width=0.8\textwidth,height=0.6\textwidth]{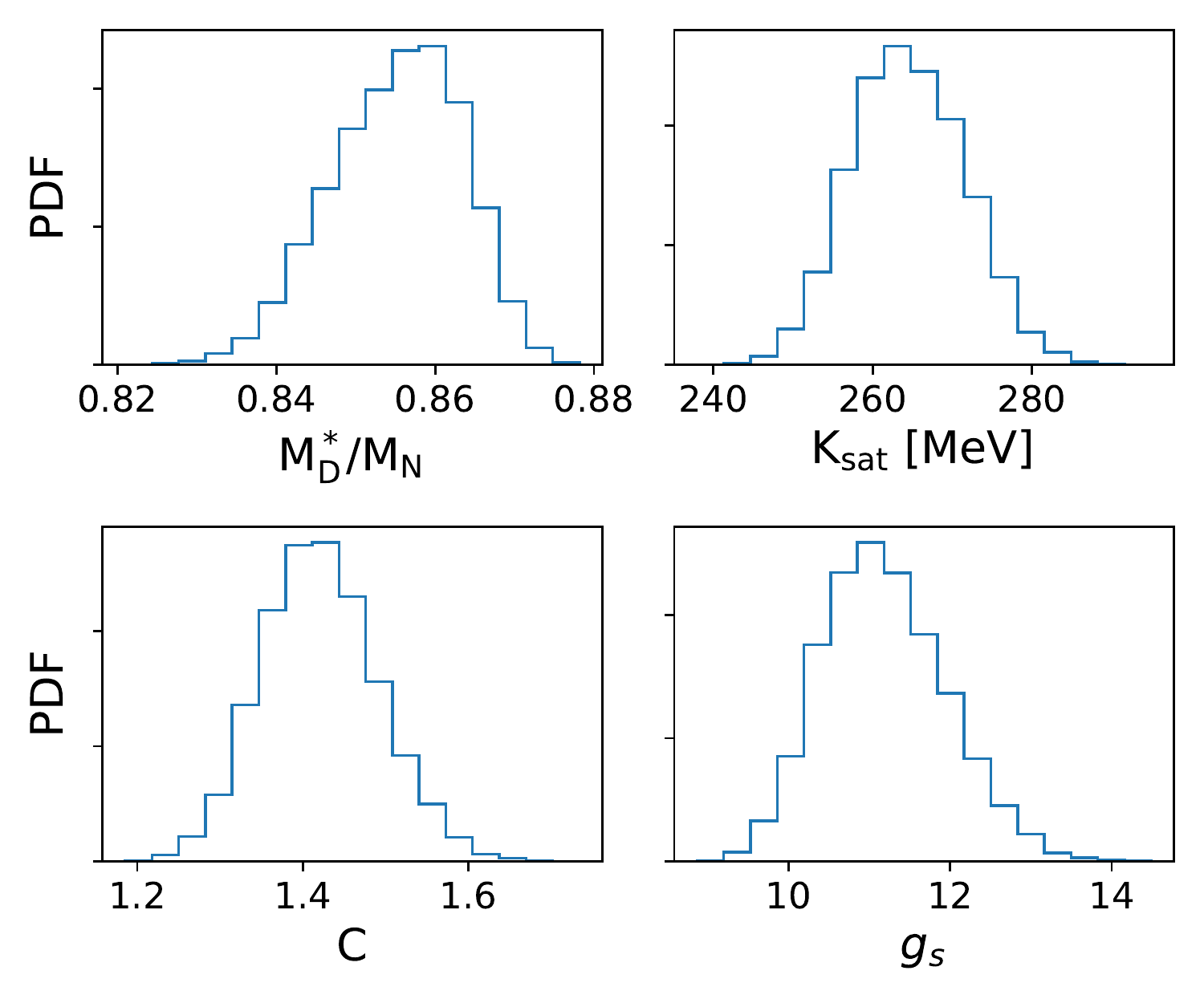}
    \caption{Prediction of the Dirac effective mass and the incompressibility modulus obtained from the RMF-CC model. The PDFs of $C$ and $g_s$ are also shown.}
    \label{fig:LSM_M_Dirac}
\end{figure*}

We will now make the connection with Lattice QCD (L-QCD) more clear by following the approach of Refs. \cite{Chanfray2007,Massot2008}. The structure of the nucleon, in particular its mass can be obtained from L-QCD, see for instance Ref.~\cite{CP-PACS:2001vqx}. In this reference precise calculations were limited to quark masses much larger than the physical one. Therefore, extrapolation of L-QCD results to the physical value of the quark mass is required, but such extrapolations run into diffcutly due to the fact that $M_N$ is a non-analytic function of $m_q$ (or equivalently $m_\pi^2$). Such non-analytic behaviour arises due to contributions from pion loops. Following the strategy of Refs.~\cite{LWY2003,TGLY04}, we express the nucleon mass as
\begin{equation}
    M_N(m_\pi^2) = a_0 + a_2 m_\pi^2 + a_4 m_\pi^4  + \dots + \Sigma_\pi,
    \label{eq:lattice_1}
\end{equation}
where we have isolated two contributions: one analytic in $m_\pi^2$ (the terms before the dots) and another, containing a non-analytic piece, denoted by $\Sigma_\pi$ which is identified with the pion self energy contribution to the nucleon mass. Note that in reality Eq.~\eqref{eq:lattice_1} is an expansion in the quark mass $m_q$, but we have replaced the quark mass with the pion mass (squared) using the GOR relation $m_\pi^2 \propto m_q$~\cite{Massot2008}. The derivative of Eq.~\eqref{eq:lattice_1} with respect to the pion mass gives the so-called sigma commutator $\sigma_N$, i.e.
\begin{equation}
    \sigma_N \equiv  m_\pi^2 \frac{dM_N}{d(m_\pi^2)} = a_2 m_\pi^2 + 2 a_4 m_\pi^4  + \dots + m_\pi^2 \frac{d\Sigma_\pi}{d(m_\pi^2)}.
\end{equation}

The sigma commutator $\sigma_N$ is an important quantity because on the one hand -- on the theory side -- it is related to symmetry properties and their explicit breaking and, on the other hand, it can be extracted from experimental results. It can also be calculated via L-QCD~\cite{Chanfray2011}.
The authors of Refs.~\cite{LWY2003,TGLY04} 
showed that it is possible to estimate the non-analytic pion self energy contribution $\Sigma_\pi$ and its derivative in an essentially model independent way using chiral perturbation theory, with the pion loops suitably regularized. Then, the parameters $a_2$ and $a_4$  are fit to L-QCD results~\cite{CP-PACS:2001vqx} and one obtains a range of values for $a_2$ and $a_4$, given in Table.~\ref{tab:param}, due to the ambiguity in the regulator of the pion loops, for which four different functional forms are used: sharp-cutoff, monopole, dipole and Gaussian \cite{TGLY04}.

Furthermore, the parameters $a_2$ and $a_4$, which are related to the analytic, non-pionic piece of $\sigma_N$,  can be used to determine the parameters of the RMF-CC model: $g_s$, $m_\sigma$ and $C$ (see Refs.~\cite{Chanfray2007,Massot2008}) using the relations
\begin{equation}
\label{eq:qcd1}
    a_2 = \frac{g_s f_\pi}{m_\sigma^2},
\end{equation}
and
\begin{equation}
\label{eq:qcd2}
    a_4 = -\frac{f_\pi g_s}{2m_\sigma^4} \bigg(3-2C\frac{M_N}{f_\pi g_s} \bigg).
\end{equation}
Notice that in the expression of $a_4$  the factor $M_N/f_\pi g_s$ was absent in \cite{Chanfray2007,Massot2008} since the nucleon mass was fixed to be $M_N=f_\pi g_s$.

In our approach we therefore have four fit parameters $a_2$, $a_4$, $m_\sigma$ and $g_\omega$. These are fixed by the analysis of the Lattice results and by two saturation properties, $n_{sat}$ and $E_{sat}$ (Table.~\ref{tab:param}). Considering the uncertainties in these parameters, one can also predict the Probability Distribution Functions (PDF) for the parameters: $g_\sigma \equiv g_s$ and $C$ from Eqs.~\eqref{eq:qcd1} and \eqref{eq:qcd2}, as well as $K_\sat$ and the Dirac mass $M_D^*$. 
Note that in SM, at the Hartree approximation, the Dirac mass is the same as the s-field dependent nucleon mass, i.e. $M^*_D = M_N(s)$.
The uncertainties in the quantities to fit (given in Table.~\ref{tab:param}) are explored within a Bayesian method using Markov-Chain Monte-Carlo (MCMC) approach. In this way full exploration of the uncertainties in the Nuclear Empirical Parameters (NEP) and the Lattice parameters are translated into uncertainty in the model parameters. The PDFs obtained from the MCMC sampling over our fit parameters $a_2$, $a_4$, $g_\omega$ and $m_\sigma$ are shown in Fig.~\ref{fig:LSM}. The distributions over $a_2$ and $a_4$ are almost flat (as imposed in the prior), and the confrontation against the NEP changes very little. 
The distributions over $g_\omega$ and $m_\sigma$ are much more peaked. The PDFs of $g_\omega$ is peaked around $6.5$. 
The PDF of $m_\sigma$ is peaked around $820$~MeV which is larger than the scalar mass ($\approx 500-600$~MeV) usually considered by RMF approaches.
The mass $m_\sigma$ is indeed related to the physical origin of the "nuclear physics" sigma meson, which is still a controversial subject since there is no sharp scalar resonance observed in the expected mass range $\approx 600$~MeV. 
Instead a broad resonance, usually refereed as $f_0(600)$, observed at around 600~MeV, is a $\pi\pi$ resonance which has no direct relation with the background scalar field introduced above.

Let us discuss this controversy in some detail, by repeating arguments already presented in Ref.~\cite{Chanfray2011}. 
The emergence of a scalar field is linked to the presence of a quark condensate, see for instance the Nambu-Jona-Lasinio model (NJL) which describes the chiral symmetry breaking in the QCD vacuum.
This scalar field is by construction a low momentum concept which does not imply the existence of a sharp scalar meson: it has been demonstrated by Celenza et al \cite{CSWSX95,CWS01} that the inclusion of a confining interaction on top of the NJL model pushes the $q\bar q$ scalar state, located originally at twice the constituent quark mass, well above one GeV. 
Coming back to the nuclear physics $\sigma_W$ and its associated ``scalar mass'', it is indeed a low momentum parameter related to the inverse of the vacuum scalar susceptibility, typically of the order of 600 to 800~MeV. There is then no need to associate it to any existing meson in the physical world.

\begin{figure*}
    \centering
    \includegraphics[width=0.8\textwidth,height=0.6\textwidth]{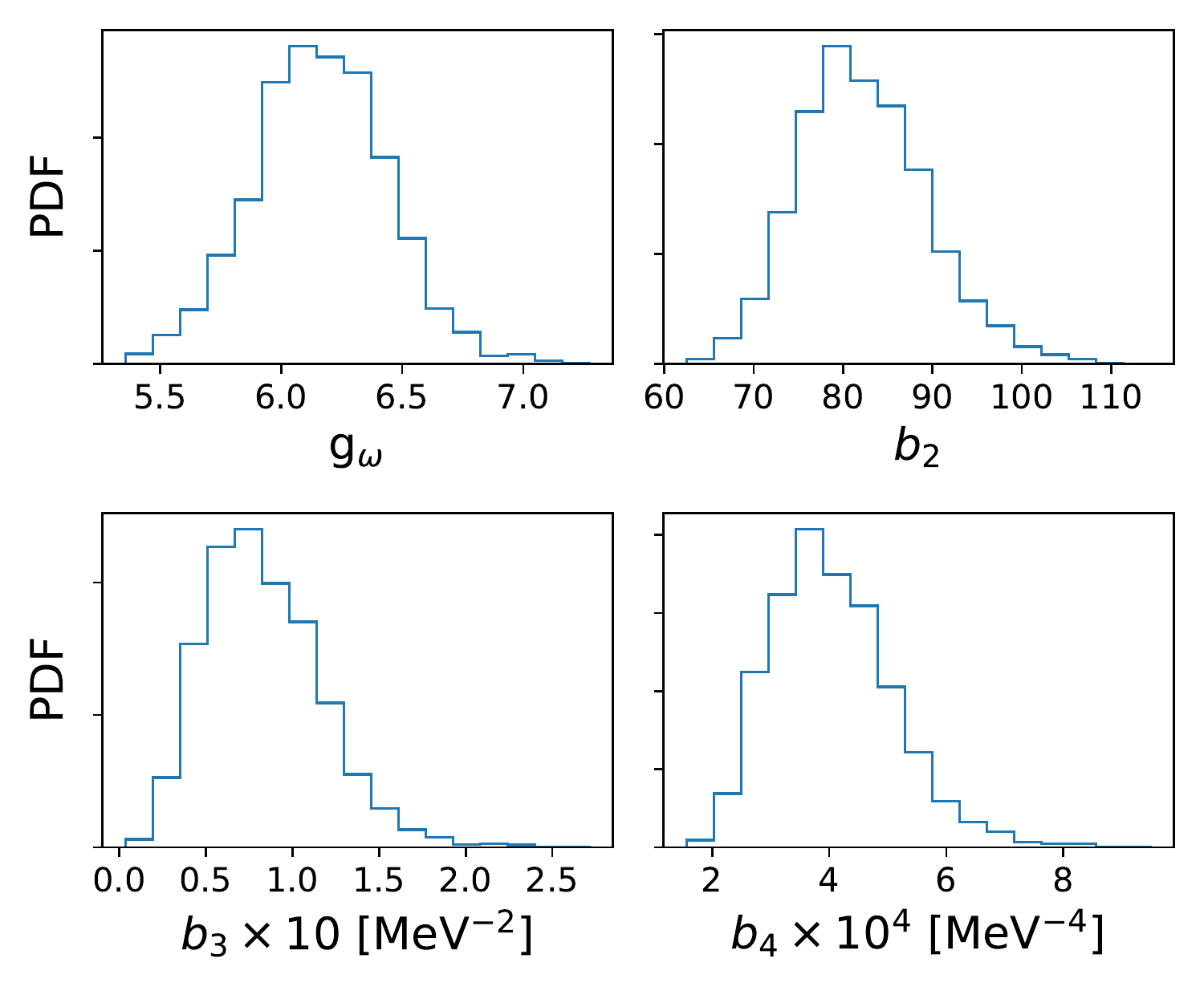}
    \caption{Results from the fit of the RMF-C model to the NEPs of Table~\ref{tab:param} and the RMF-CC model's prediction of $M^*_D$ and $K_{\sat}$ shown in Fig.~\ref{fig:LSM_M_Dirac}.}
    \label{fig:chiral_pasta}
\end{figure*}

Based on the parameter distributions shown in Fig.~\ref{fig:LSM}, we can now analyze the impact of the uncertainties in these parameters on several interesting properties of dense matter, e.g. the Dirac mass at saturation $M^*_D$ and the incompressibility modulus $K_{\sat}$. For completeness, we also show the distribution over the parameters $C$ and $g_\sigma$. These results are shown in Fig.~\ref{fig:LSM_M_Dirac}. The Dirac mass is peaked around $\approx 0.85\pm0.02 M_N$.
The predictions for $K_{\sat}$, with a PDF peaked at $\approx 265$ MeV, are slightly larger than the expected empirical value around $230-250$~MeV~\cite{Margueron2018}. We expect however that quantum corrections, e.g. Fock term or pion cloud~\cite{Massot2009}, could change these quantities and shift them towards lower values. The PDF of $C$ is consistent with the value used in Ref.~\cite{Massot2008} and $g_s$ is consistent with the canonical value of $M_N/f_\pi \approx 9.98$.

Note that the nucleon response $\kappa_{NS}$ contributes as well to the curvature coefficient at saturation -- the incompressibility modulus $K_\sat$. In a set of successive works \cite{Chanfray2005,Chanfray2007,Massot2008,Massot2009,Chanfray2011,Massot2012} this approach has been applied  to the equation of state of nuclear matter and neutron stars as well as to the study of chiral properties of nuclear matter at different levels of approximation in the treatment of the many-body problem (RMF, Relativistic Hartree Fock or RHF, pion loop correlation energy). Note also that, the quark substructure plays also a crucial role for the spin-orbit potential as discussed in a recent paper~\cite{CM2020}.

\subsection{Relativistic Mean Field with Chiral Symmetry only (RMF-C)}

We now consider an approach where chiral symmetry is incorporated within a chiral potential $V(s)$, but without the effect of confinement in terms of nucleon polarisation. This so-called RMF-C model is inspired from Refs.~\cite{Wetterich,Drews2013,Fraga2018,Schmitt2020}. 

Several chiral relativistic theories (of the RMF-C type) have indeed been formulated but without reference to the nucleon response \cite{Boguta83,Boguta89,Wetterich,Drews2013}. Recently, such a RMF-C model has been used to study the possible mixed phase at the chiral transition in SM~\cite{Fraga2018} and neutron stars \cite{Schmitt2020}. Note that our chiral invariant $S$ field is named $\chi$ in the latter paper. In this approach the chiral potential deviates from the pure linear sigma model potential (used by our RMF-CC model) by terms of first and third order in  $S^2-f^2_\pi$ with additional parameters ($a_3$, $a_4$). This is a legitimate attitude since any microscopic underlying model, including the NJL model for instance, will certainly generate such higher order many-body terms at low-energy. 

Since the non-trivial scalar response of the nucleon is neglected in this model, the $s$-field dependent nucleon mass is simply given by
\begin{equation}
M_N(s) = M_N + g_s s \,,
\end{equation}
as in the linear sigma model~\cite{Bhaduri1988}.

\begin{figure*}
    \centering
    \includegraphics[width=0.8\textwidth,height=0.6\textwidth]{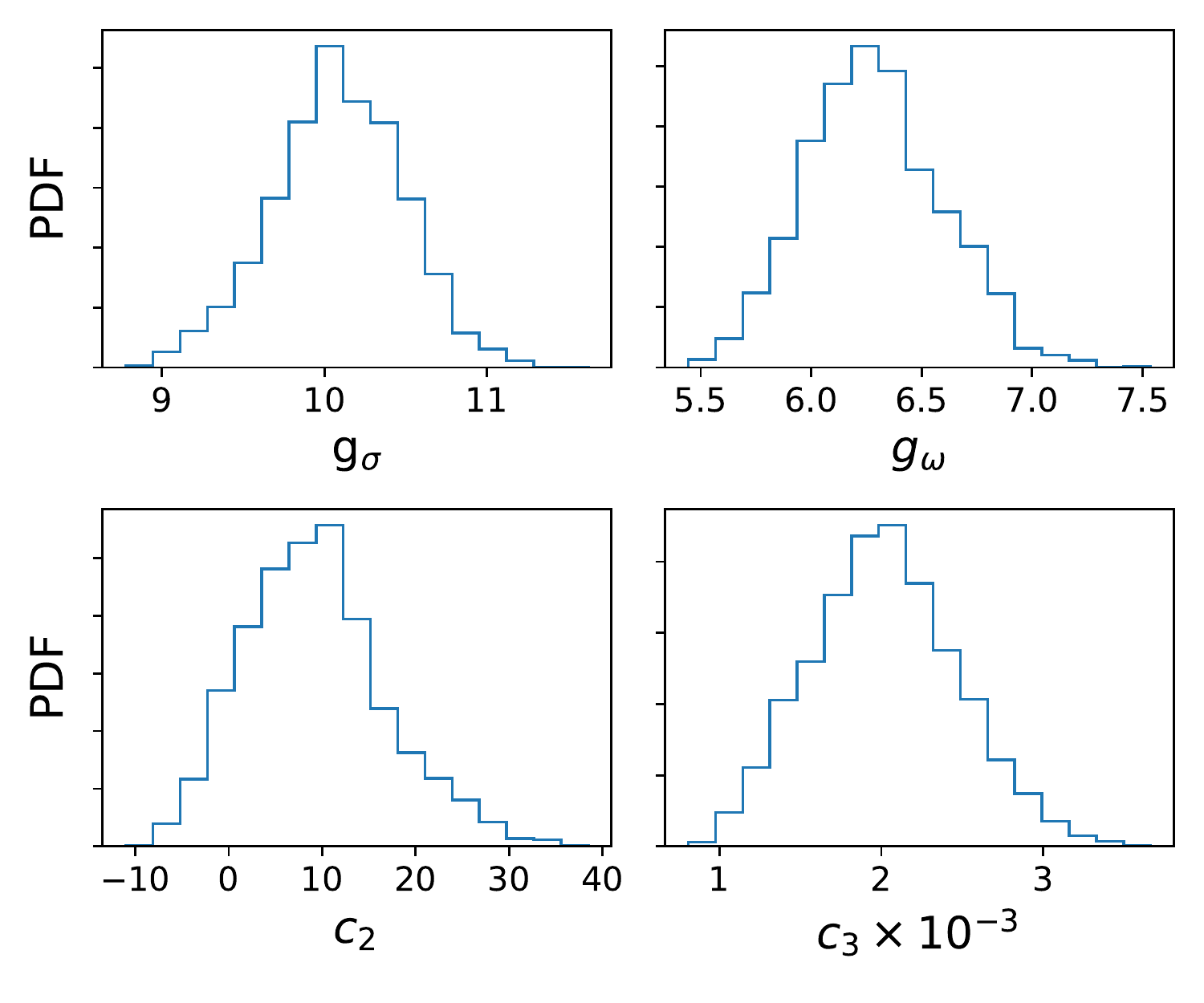}
    \caption{Results from the fit of the RMF model to the NEPs of Table~\ref{tab:param} and the RMF-CC model's prediction of $M^*_D$ and $K_{\sat}$ shown in Fig.~\ref{fig:LSM_M_Dirac}.}
    \label{fig:RMF}
\end{figure*}

The chiral potential is expressed in a very general way as
\begin{eqnarray}
\label{eq:V_pasta}
V(s) = \sum_{n=1}^4 \frac{b_n}{n!} \frac{(S^2 - f_{\pi}^2)^n}{2^n} - f_{\pi} m_{\pi}^2 \sigma.
\end{eqnarray}
Notice that contrary to Eq.~\eqref{eq:V(s)_exp}, this chiral potential contains terms of order $n=3,4$ in the $S^2-f_\pi^2$ expansion.

One interesting question is whether this higher order terms may simulate the effect of the nucleon response. Since a term in $s^n$ corresponds to a $(n-1)$-body force, one may thus expect that the expansion in the scalar field $s$ is perturbative, at least at low density: the 2-body force is expected to be larger than the 3-body one, itself larger than the 4-body force, and so on. Models violating this ordering will thus be referred as anomalous ones in the following.
Anticipating our results, we found that this RMF-C model lead to anomalous chiral potentials.

An important specific point is that, as in the linear sigma  model, the scalar coupling parameter is fixed, $g_s = M_N/f_\pi$. This leaves us with 4 unknown parameters $g_\omega$, $b_2$, $b_3$ and $b_4$ that we fit in a consistent way compared to RMF-CC.
To do so, we consider the two NEPs in Table~\ref{tab:param}, $E_\sat$ and $n_\sat$ as in the RMF-CC model and, additionally we use the predictions of the RMF-CC model for the Dirac effective mass and the incompressibility modulus shown in Fig.~\ref{fig:LSM_M_Dirac}. 

The result of sampling this distribution in the parameter space spanned by $g_\omega$, $b_2$, $b_3$ and $b_4$ is shown in Fig.~\ref{fig:chiral_pasta}. 
First let us comment on the marginalized 1-dimensional PDF over $g_\omega$. The distribution is peaked around $\approx 6.25$. This is a rather low value given that the value of $g_\omega$ found in Ref.~\cite{Schmitt2020} is $9.47$. This discrepancy is due to the fact that we have fitted $g_\omega$ to a Dirac effective mass $M^*_D$ that is $\approx 0.85 M_N$ (see Fig.~\ref{fig:LSM_M_Dirac}), whereas the value of $M^*_D$ used in Ref.~\cite{Schmitt2020} is $0.75M_N$.
We have verified that if $M^*_D = 0.75M_N$ is used in our approach, we are able to reproduce the $g_\omega$ found in Ref.~\cite{Schmitt2020}.
Regarding the other parameters $b_2$, $b_3$ and $b_4$, the PDFs are also peaked at values different from the ones found in Ref.~\cite{Schmitt2020}. 
This is due to the fact the NEP used in this work are different from that used in Ref.~\cite{Schmitt2020}, most notably $K_{\sat}$
but also $E_{\sat}$ and $n_{\sat}$, therefore a precise agreement of our results should not be expected. Indeed for this work what is important is that the three models are parametrized consistently at saturation density. 
In this way, the differences in the predictions at high density could only be related to the ingredients of the models.

Finally, let us note that the result of $g_\omega$ presented here is consistent with what one obtains in the RMF-CC model, see Fig.~\ref{fig:LSM}. We also remark that the PDFs are broad and thus the parameters cannot be very well constrained by empirical knowledge of SM saturation. We will comment more on the values of these parameters later when comparing the scalar potentials of different models.

\subsection{Relativistic Mean Field Theory (RMF)}

We now turn to a model in which both the chiral potential and the response of the nucleon is ignored. 
The first attempt to go beyond a non relativistic treatment of nuclear matter was the relativistic mean field (RMF) approach initiated  by Walecka and collaborators \cite{SerotWalecka1986,Walecka1997}, which is based on meson exchange between nucleons whose wave functions are solution of the in-medium Dirac equation. In this framework nucleons move in an attractive (scalar field) and in a repulsive (vector field) backgrounds. This provides both the "Walecka" saturation mechanism and the correct magnitude of the spin-orbit potential. The parameters describing the meson-nucleon couplings are adjusted to the saturation properties of nuclear matter and/or nuclear ground state properties through the nuclear chart (binding energies, charge radii, etc). Hence there is no explicit or direct connection with the underlying QCD theory but instead this approach  describes the nuclear properties in terms of a meson exchange potential renormalized around nuclear saturation density. 

As in the original Walecka model, the scalar potential is limited to the mass term:
\begin{eqnarray}
V(\sigma_W) = \frac{1}{2}m_\sigma^2 \sigma_{W}^2 \, ,
\label{eq:walecka:pot}
\end{eqnarray}
and for the $\sigma_W$-field dependent mass one has:
\begin{eqnarray}
M_N(\sigma_W) = M_N + g_\sigma \sigma_W\, ,
\label{eq:walecka:ms}
\end{eqnarray}
where $g_\sigma$ is the scalar coupling constant. 

While the saturation mechanism arises from the equilibrium between the scalar and the vector fields and allows a good reproduction of the saturation density and binding energy -- at the cost of large coupling constants -- other properties of the model, e.g. the compression modulus, the effective nucleon mass and the symmetry energy, are in poor agreement with the empirical values. 
Boguta and Bodmer~\cite{Boguta1977} have thus suggested an extension of the original Walecka model, whose main purpose is to bring the compression modulus and nucleon effective mass at saturation under control, by introducing self-interactions of the scalar field by modifying the potential~\eqref{eq:walecka:pot} as,
\begin{eqnarray}
V(\sigma_W) = \frac{1}{2}m_\sigma^2 \sigma_W^2 + \frac 1 3 c_2 M_N  \sigma_W^3 + \frac 1 4 c_3 \sigma_W^4.
\label{eq:pot:selfint}
\end{eqnarray}
Such self-interacting scalar field potentials have been largely employed in what is commonly referred as the Relativistic Mean-Field Model (RMF), e.g. NL3~\cite{Lalazissis1996} and see also the Glendenning book~\cite{Glendenning1997}.
Note that other extensions based on density dependent coupling constant will not be considered in the present study.
It is interesting to remark that the Euler-Lagrange equation for the scalar field is modified by the self-interaction terms, but the nucleon effective mass remains described by Eq.~\eqref{eq:walecka:ms}, as in the original Walecka model. This also makes the RMF model qualitatively similar to the RMF-C one, as we will illustrate it in the next section.

We have 4 parameters to fit, $g_\sigma$, $g_\omega$, $c_2$ and $c_3$. As with the case of the RMF-C model, we use the two NEPs of Table~\ref{tab:param} and the PDFs of $M^*_D$ and $K_{\sat}$ shown in Fig.~\ref{fig:LSM_M_Dirac}. The results of sampling this distribution is shown in Fig.~\ref{fig:RMF}. We see that the PDF of $g_\sigma$ is peaked around $10$, which is consistent with what is shown in Fig.~\ref{fig:LSM_M_Dirac} for the RMF-CC model. Note that we fix $m_\sigma = 800 \MeV$ compatible with the peak predicted by the RMF-CC model. Fixing the value of $m_\sigma$ in RMF is not constraining if $g_\sigma$ is varied: only the ratio $g_\sigma/m_\sigma$ matters. We remind that in the RMF-CC model however this degeneracy is broken by Eqs.~\eqref{eq:qcd1} and \eqref{eq:qcd2}. Additionally, the fixing of $m_\sigma$ at a constant value can be seen as analogous to the RMF-C model where the parameter $g_s$ is frozen instead of $m_\sigma$.
For $g_\omega$ we again obtain a value $\approx 6.25$ which is again due to the fact that we fit to a large value of $M^*_D \approx 0.85 M_N$. This value of $g_\omega$ is very close to those obtained previously for the RMF-CC and RMF-C models. 
Finally, we note that the PDF of the $c_2$ parameter has a peak around 10 but with an uncertainty of about 10, making it compatible with 0. The PDF of $c_3$ prefers very large values. We will comment extensively on our results for $c_2$ and $c_3$ later when we compare the scalar potentials of the different models. 

\begin{figure}
    \centering
    \includegraphics[width=0.45\textwidth,height=0.4\textwidth]{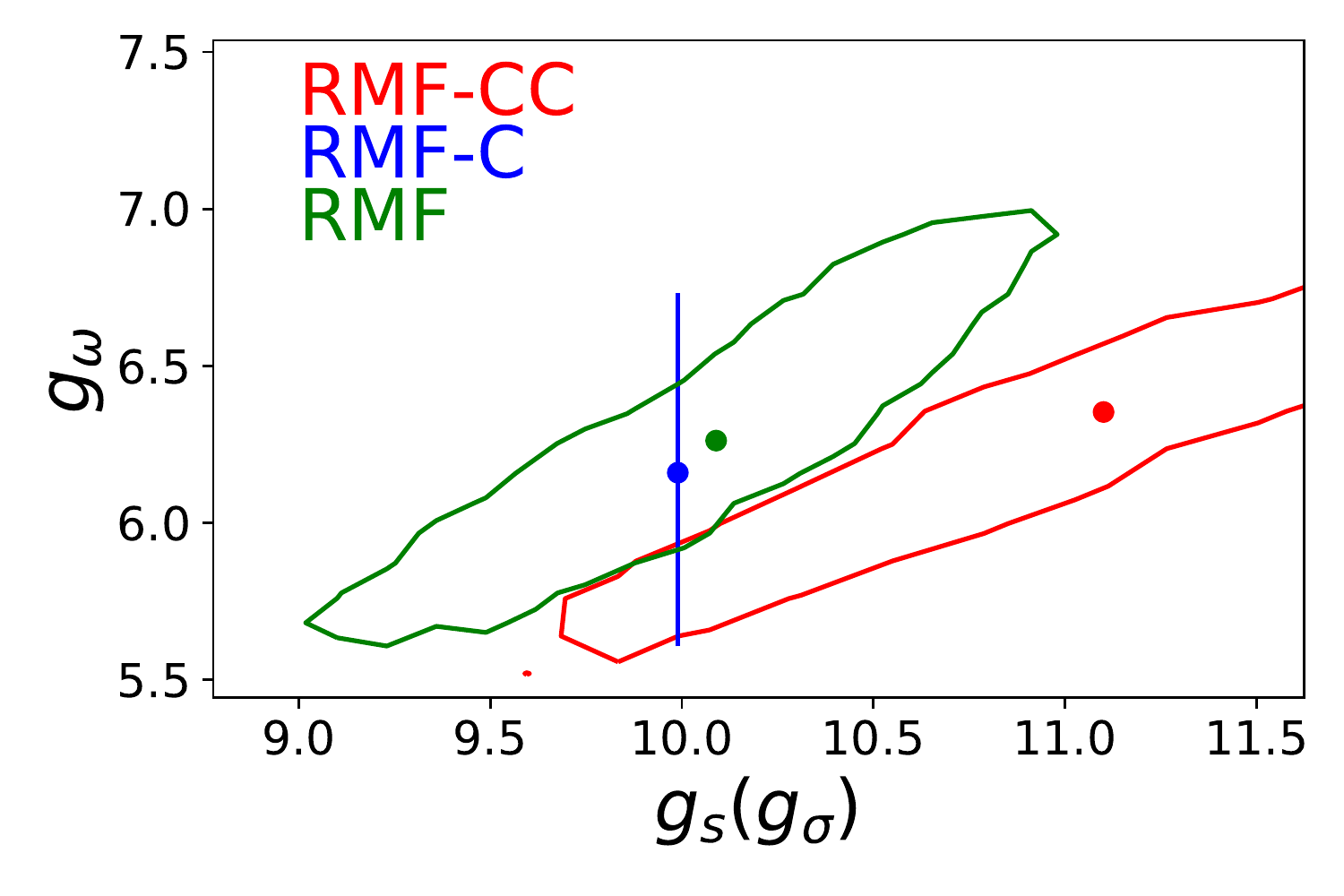}
    \caption{The correlation between $g_s$ and $g_{\omega}$ for the three models. The 95 \% confidence levels are shown. The dots represent the centroids of the distributions.}
    \label{fig:scatter}
\end{figure}

Having obtained the values of $g_s (g_\sigma)$ and $g_\omega$ for the three models, we can study how the three models can be separated when the correlation between $g_\omega$ and $g_s$ is analysed. In Fig.~\ref{fig:scatter}, this correlation is plotted for the three models in different colors, where the contours represent the 95\% confidence level. For the RMF-C model, only a vertical line is shown since $g_s$ is fixed in this case. We see that for the three models, the centroid of $g_\omega$ are very close ($\approx 6.25$). However, the models can be separated along the horizontal coordinate ($g_s$). The RMF-CC model prefers larger values of $g_s$, whereas the RMF-C and RMF models prefer the lower value close to $M_N/f_\pi$.

\section{Comparison of the three classes of relativistic models}
\label{sec:comaparison}

In the previous section, the three models have been fit to reproduce the same properties at saturation in SM. These properties are the saturation density and energy for all models, and the models RMF-C and RMF are adjusted to reproduce the same Dirac mass and incompressibility modulus as RMF-CC, which, for this model are deduced from fundamental L-QCD properties.
The models are therefore treated on an equal footing by ensuring that they agree on the empirical parameters and their uncertainties: $n_{\sat}$, $E_{\sat}$, $K_{\sat}$ and $M^*_D (n=n_{\sat})$. 

In this section we will show that although the predictions of these three models agree at saturation density, they 
differ quantitatively at larger densities since they represent different density functionals. Moreover, a detailed analysis of the scalar field properties indicates that RMF-CC represents a microscopically justified and an economical way to incorporate in-medium corrections on top of the chiral potential defined in the vacuum.

\subsection{The energy per particle, the self-energies and the effective masses}
\label{sec:energy}

\begin{figure*}
    \centering
    \includegraphics[width=0.33\textwidth]{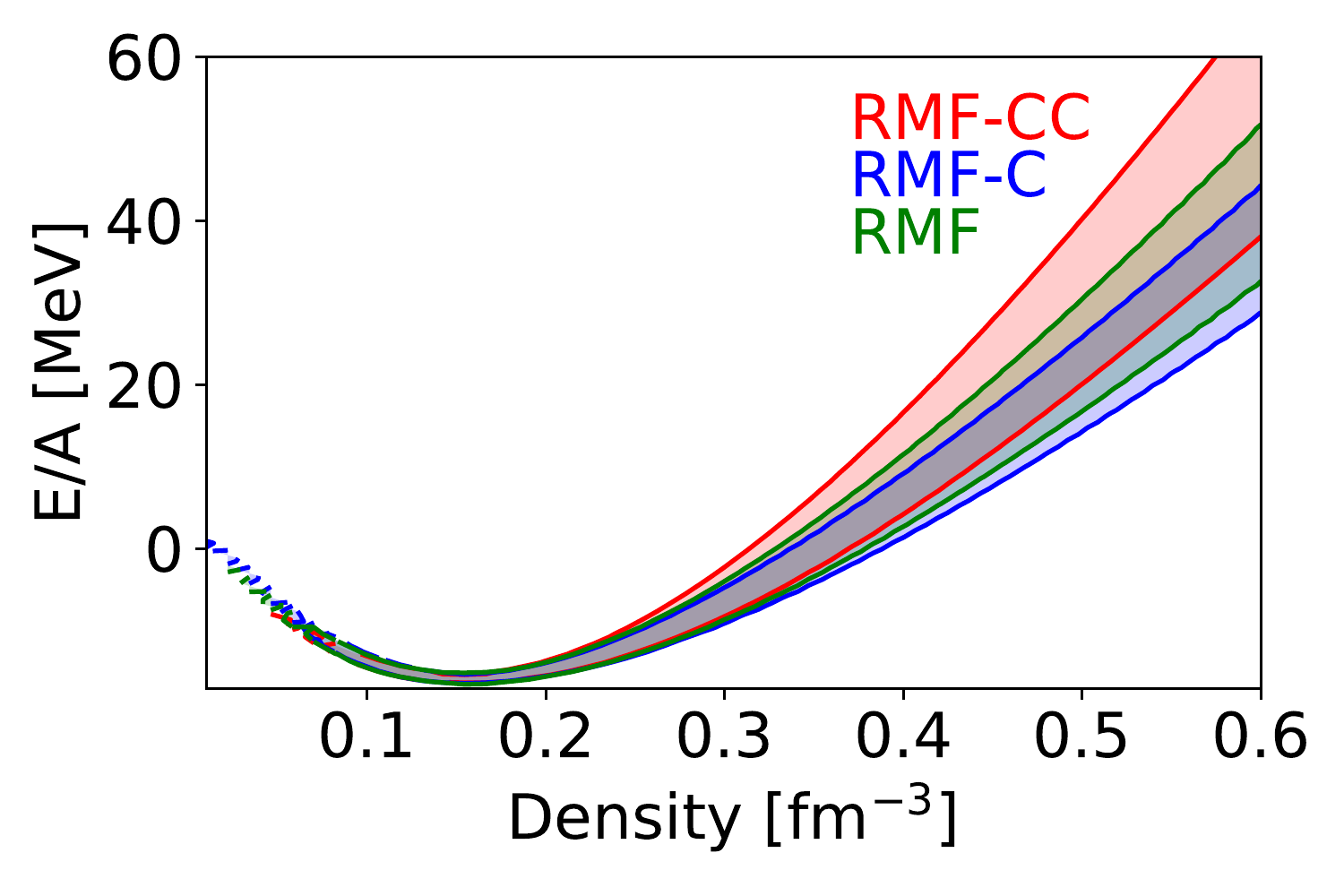}
    \includegraphics[width=0.32\textwidth]{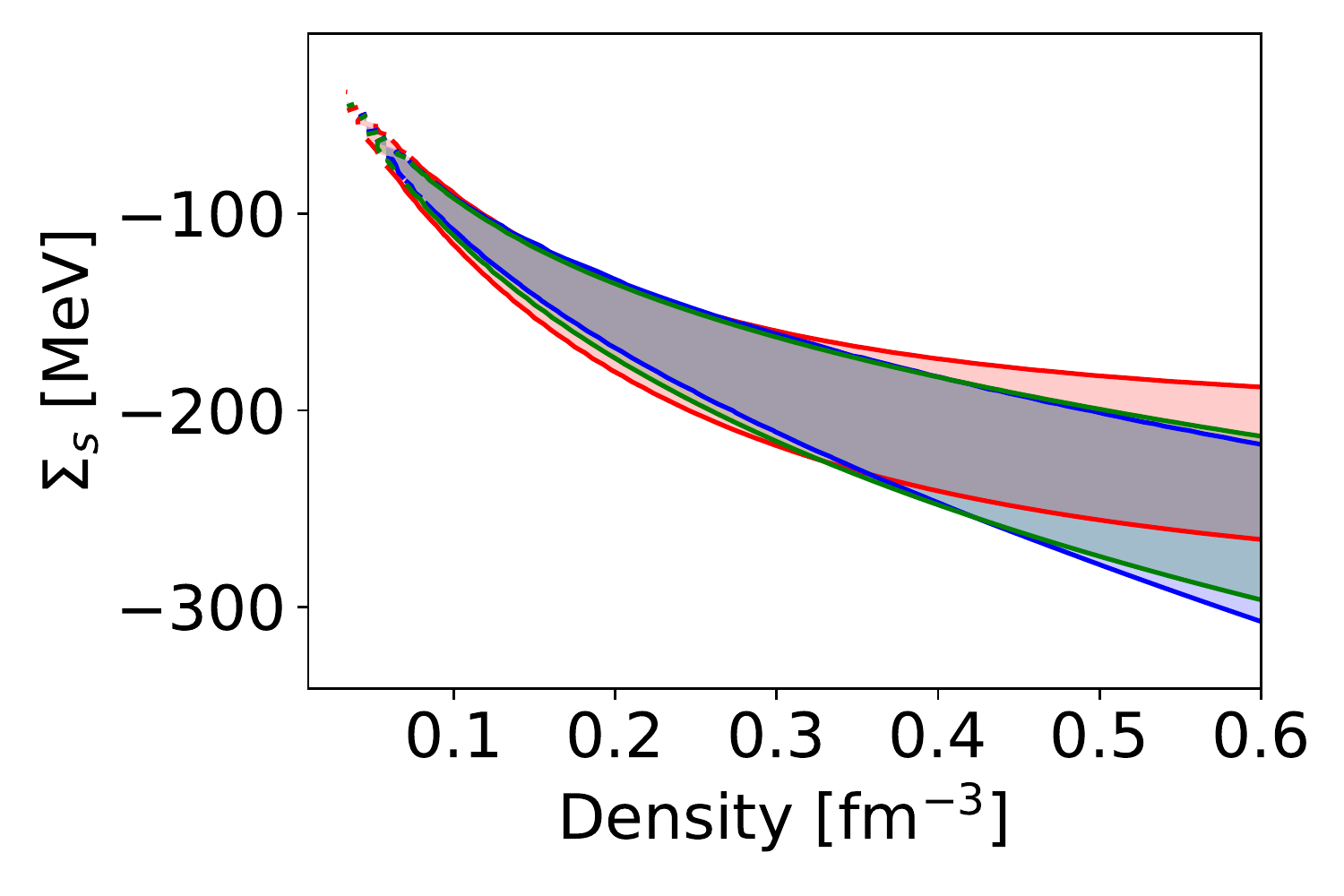}
    \includegraphics[width=0.32\textwidth]{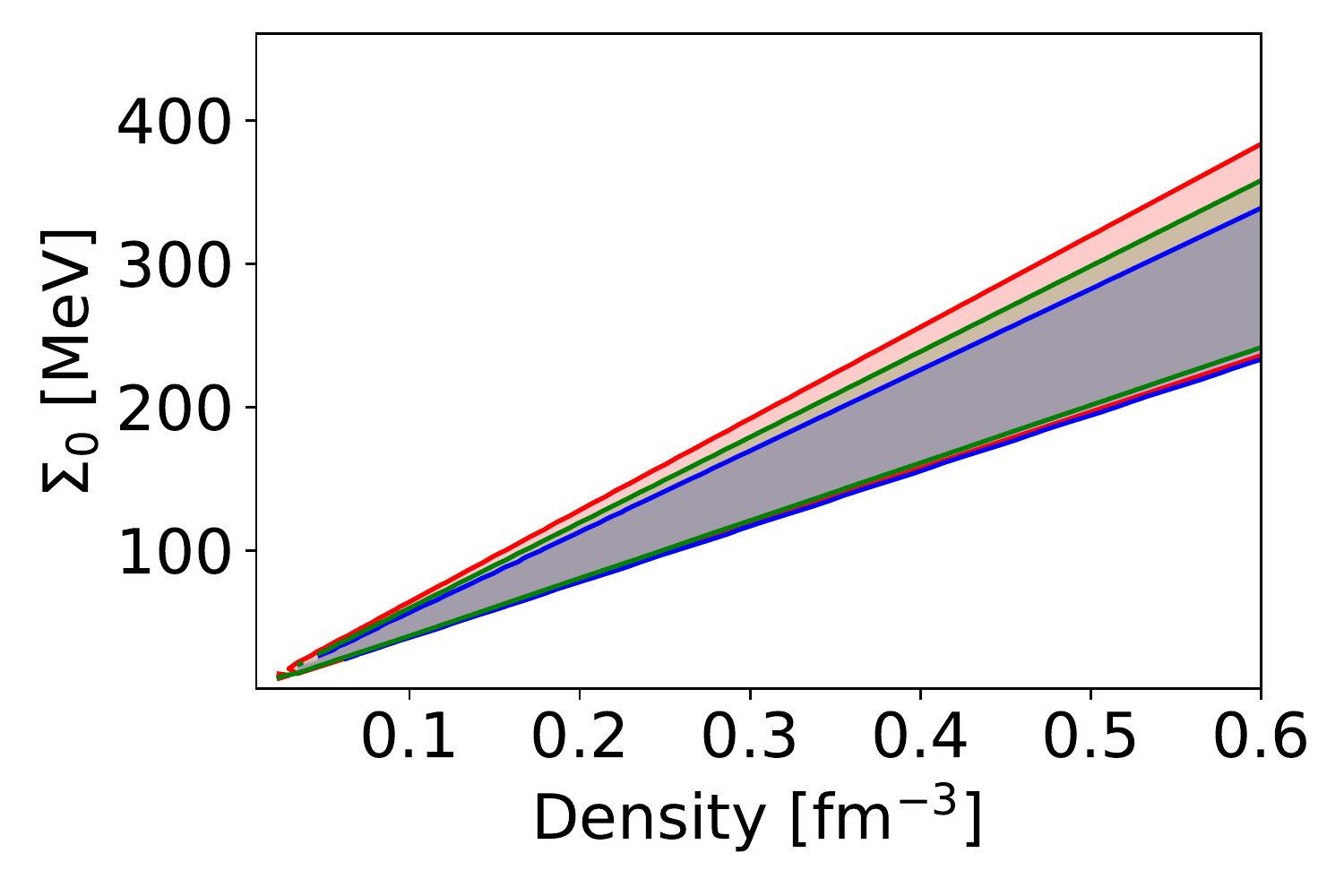}
    \caption{(left) The energy per particle in SM for the three models considered in this work. The contours show the 95\% confidence level.
    The density dependence of the scalar self energy (center panel) and the time component of the vector self energy (right panel) are also shown.} 
    \label{fig:e}
\end{figure*}

\begin{figure}
    \centering
    \includegraphics[width=0.45\textwidth]{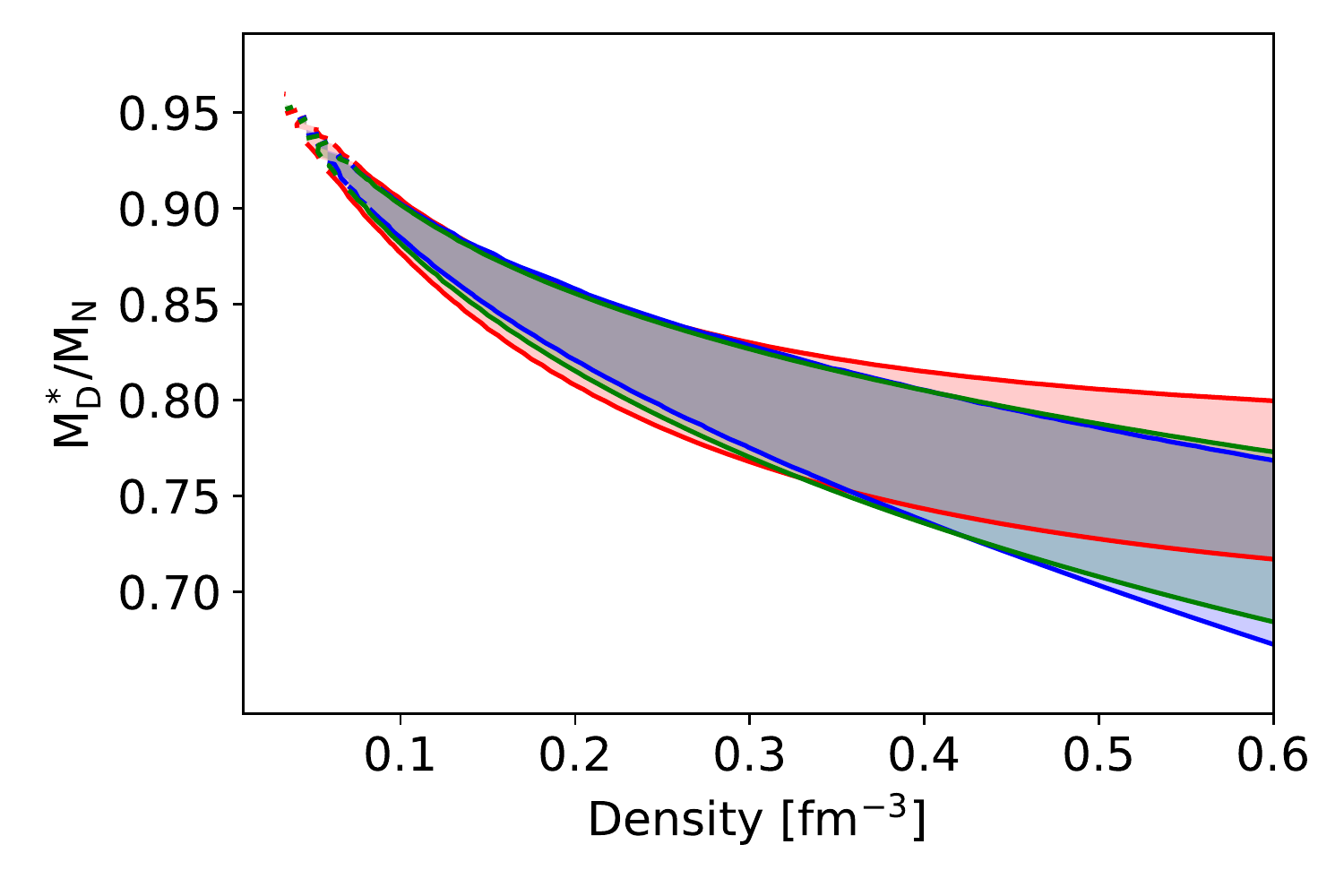}
    \includegraphics[width=0.45\textwidth]{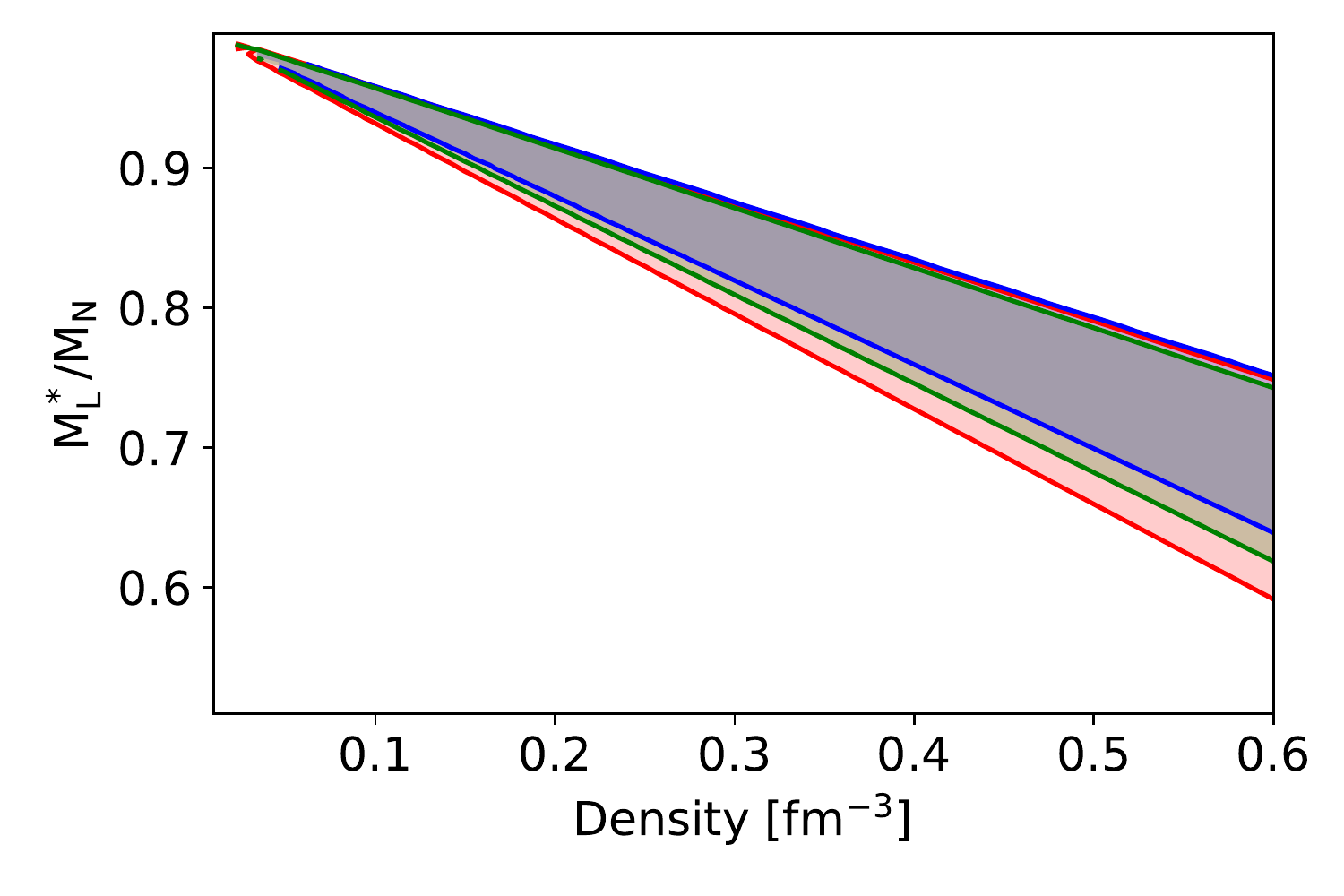}
    \caption{The density dependence of the Dirac and Landau masses are shown.} 
    \label{fig:eff_mass}
\end{figure}

We first start with an analysis of the energy per particle in SM. In Fig.~\ref{fig:e}, the results are shown in the left panel. The three models correspond to the three colors. 
The upper and lower limits represent the 95\% CL, allowing a visualization of the uncertainties in the model predictions as a function of density. Recall that these uncertainties originate from our imperfect knowledge of nuclear matter saturation properties and fundamental predictions of L-QCD.   
We see that the three models agree well at densities $n \approx n_{\sat}$, since they are constrained to do so. The agreement also appears to be quite good at $n < n_{\sat}$. However, at $n > 2 n_{\sat}$, RMF-CC model predicts the larger values for the energy per particle, while RMF-C model produces the smaller ones and RMF model lies in the intermediate range. Note however, that all model predictions are compatible with each other within the considered 95\% confidence levels. Given that $g_\omega$ is similar for all three models, the reason for differences at high densities is most probably related to the scalar interaction, i.e. the scalar potential and/or the scalar coupling to the nucleons. In the next section, we will investigate the former in detail.  

In the center and right panels of Fig.~\ref{fig:e}, the scalar and the vector (time component) self energies, $\Sigma_s$ and $\Sigma_0$ are shown. At the mean-field level in SM, we have
\begin{eqnarray}
    \Sigma_s &=& M_N(s) - M_N \, \\
    \Sigma_0 &=& g_\omega^2/m_\omega^2 \rho  \, .
\end{eqnarray}
We see again that the three models agree at low densities. At larger densities RMF-CC predicts a slightly larger value for $\Sigma_s$, however there is still significant overlap among the predictions. For $\Sigma_0$, the models still agree at large densities. This is expected since $g_\omega$ is the only parameter that controls the density dependence of $\Sigma_0$, and all three models have similar values of $g_\omega$. Since correlations beyond the mean field lead to a more complicated density dependence of $\Sigma_0$~\cite{Ma2004,Jaminon1989}, it would be interesting to re-analyse this quantity by including Fock contributions in the future. 

Finally, in Fig.~\ref{fig:eff_mass}, the Dirac and Landau masses ($M^*_D$ and $M^*_L$) are shown. The Landau mass has been computed by deriving the Schrödinger equivalent single-particle potential following Refs.~\cite{Ma2004,Jaminon1989}. At the Hartree approximation, it reads
\begin{equation}
    M^*_L = M_N - \Sigma_0 .
\end{equation}
On the other hand, the Dirac mass is the same as the s-field dependent nucleon mass, i.e. $M^*_D = M_N(s)$. All the comments made regarding $\Sigma_s$ and $\Sigma_0$ are applicable to the Dirac and Landau masses respectfully, since the relationship between the self energies and the effective masses is quite straightforward in the mean-field level. 

In summary, the three models presented here, while being calibrated on the same quantities at saturation, lead to slightly different predictions above saturation density: RMF-CC is more repulsive than RMF-C on the average, while RMF lies in between them.
In the following, we investigate more closely the properties of the microscopic quantities at the base of the models: the scalar potential and the self-consistent equation for the scalar field.

\subsection{Analysis of the scalar potential $V(s)$}
\label{sec:potential}

The scalar potentials $V(s)$ have different expressions in the models considered in our analysis. For an easy comparison, we recast the chiral potential $V(s)$ for RMF-CC and RMF-C into the form of the scalar potential in RMF, see Eq.~\eqref{eq:pot:selfint}.
In doing so, the chiral potential~\eqref{eq:V(s)_s} in RMF-CC leads to the following coupling constants,
\begin{eqnarray}
c_2^\textrm{RMF-CC} &=& \frac{3}{2 f_\pi M_N}(m_\sigma^2-m_\pi^2) \, \\
c_3^\textrm{RMF-CC} &=& \frac{1}{2 f_\pi^2}(m_\sigma^2-m_\pi^2)=\frac{M_N}{3f_\pi}c_2^\textrm{RMF-CC} \, ,
\end{eqnarray}
and for RMF-C the chiral potential~\eqref{eq:V_pasta} gives 
\begin{eqnarray}
c_2^\textrm{RMF-C} &=& \frac{1}{M_N} \bigg(\frac{3}{2}b_2 f_\pi + \frac{1}{2}b_3 f_\pi^3 \bigg) \, \\
c_3^\textrm{RMF-C} &=& \frac{1}{2}b_2 + b_3 f_\pi^2 + \frac{1}{6}b_4 f_\pi^4 \, .
\end{eqnarray}

\begin{table}[tb]
\centering
\setlength{\tabcolsep}{22pt}
\renewcommand{\arraystretch}{1.5}
\caption{Coefficients of the scalar potentials expressed as in Eq.~\eqref{eq:pot:selfint} for the three models considered here. The quoted uncertainties represent the 95\% CL.}
\label{tab:g}
\begin{tabular}{cccc}
\hline\hline
Model  & $c_2$ & $c_3$\\
\hline
RMF-CC  & $11.3^{+2.2}_{-1.7}$ & $37.5^{+7.3}_{-5.6}$\\
RMF-C  & $47.2^{+37.4}_{-23.5}$ & $5880^{+3870}_{-2470}$ \\
RMF  & $8.9^{+17.2}_{-12.6}$  & $2010^{+940}_{-820}$ \\
NL3\cite{Lalazissis1996}  & -29.89 &  -2.19 \\
\hline\hline
\end{tabular}
\end{table}

\begin{figure*}
    \centering
    \includegraphics[width=1.0\textwidth]{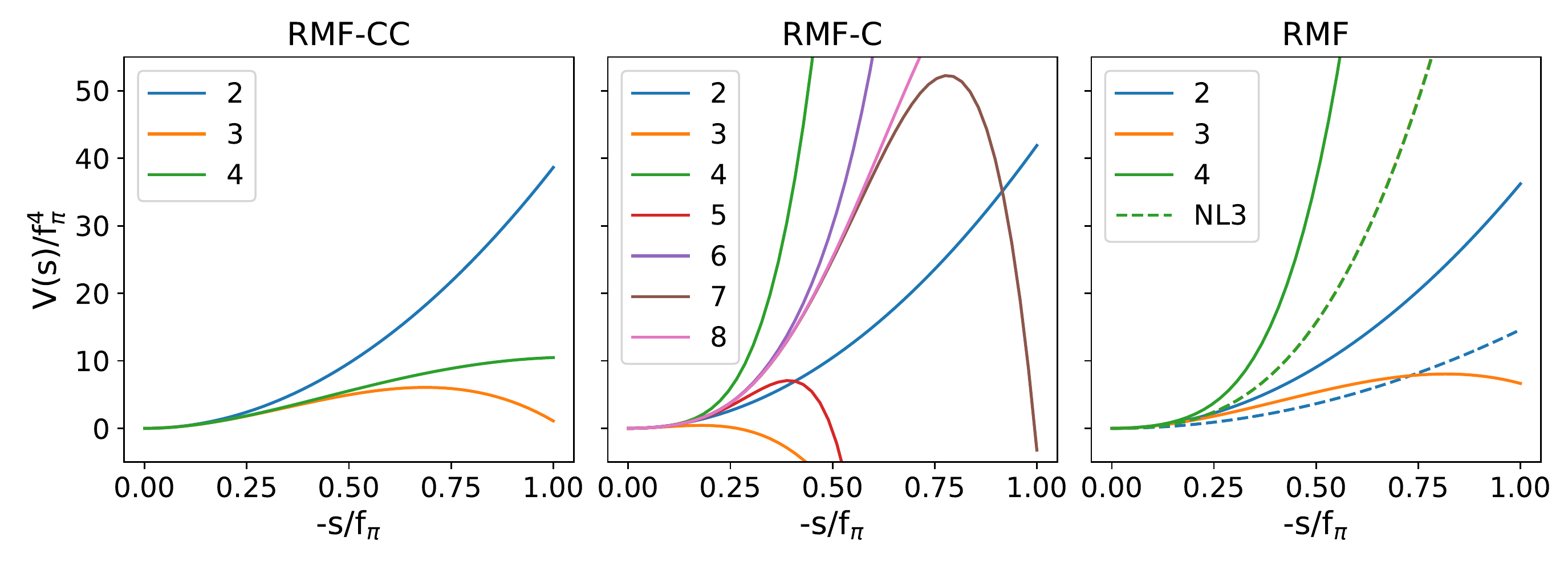}
    \caption{Analysis of the potential of the scalar field. Considering the potential as an expansion in $s$, the number in the legend refers to the order at which this expansion is truncated. The parameters of the scalar potential are taken to be the mean values reported in Table.~\ref{tab:g}. In the right panel, for the NL3 parametrization (dashed lines) the orange dashed line is hidden behind the green one.}
    \label{fig:Potential_order}
\end{figure*}

The sign of the parameters $c_2$ and $c_3$ are important for the interpretation of the scalar potential in terms of a Mexican hat potential. It indeed implies that a positive $c_2$ generates an attractive term (since $s$ is negative) and a positive $c_3$ a repulsive term. As we already pointed out, the magnitude of these parameters
is also important in order to interpret the different terms of the potential as an expansion in terms of many-body interactions, since a term in $s^n$ corresponds to an $(n-1)-$body force.
Since these many-body forces are expected to be hierarchically ordered (at least at low densities), truncation at different orders of the scalar potential are expected to evolve smoothly. When this is not the case, we will interpret it as an anomaly of the scalar potential.

In Tab.~\ref{tab:g} we compare the parameters $c_2$ and $c_3$ determined for the three models. For all the models considered in this work, except NL3, the centroids of both $c_2$ and $c_3$ are positive, as expected. The parameters $c_2$ and $c_3$ of the RMF model are found to be different from the original NL3 model of Ref.~\cite{Lalazissis1996}, where $c_2$ and $c_3$ are both negative. Since the parameters $c_2$ and $c_3$ are obtained from a fit to the NEPs, their values are determined from the values considered for these NEPs. The values for $E_{\sat}$, $n_{\sat}$, $M^*_D (n=n_{\sat})=M_N(s)$ and $K_{\sat}$ are indeed different for NL3 and the RMF case.

Large values of $c_3$ found for RMF and RMF-C indicate that $V(s)$, when considered as an expansion in $s$, might present an anomaly in the order hierarchy. To make it more clear, we show in Fig.~\ref{fig:Potential_order} the chiral potential truncated at various orders in $s$, starting from order 2. In RMF-CC, the distinction between order 3 and order 4 curves appears only at large $s$, and the 4th order correction is relatively small. In the case of RMF-C however, we see that every addition of a higher order correction drastically changes the behaviour of the scalar potential. Indeed, a truncation at order 3 or 5 would result in an overall change of sign of $V(s)$ at $s \approx 0.5$. Therefore in the case of RMF-C, the correct reproduction of nuclear NEPs in SM is due to a fine tuning between the parameters $b_2$, $b_3$ and $b_4$, rendering difficult the interpretation of $V(s)$ in terms of many-body forces. 
We thus qualify the chiral potential in RMF-C as presenting an anomaly. We have a similar behaviour for RMF. The 4th order correction to the order 3 curve is very large, which is imposed by the saturation properties of RMF-CC. The scalar potential of RMF is thus also possibly anomalous. This conclusion is of course limited to the explored parameters region considered in our study - and related to the predictions of RMF-CC model - but different parameter sets could lead to a convergent expansion, as illustrated for instance by NL3 (RMF model), see the right panel of Fig.~\ref{fig:Potential_order}. Note that in this case the 4th order correction (dashed green line) is so small that it lies on top of and thus hides the 3rd order term.

In conclusion, we observe that if the three models RMF-CC, RMF-C and RMF are constrained to reproduce the same properties at saturation, the $s$ expansion of the scalar potential $V(s)$ may manifest an anomalous behaviour for RMF-C and RMF, at variance with RMF-CC. In the following section, we show that the origin of this anomaly for RMF-C and RMF can be related to the absence of the scalar nucleon response in their Lagrangian.

\subsection{Analysis of the equation of motion of the scalar field}
\label{sec:eom}

\begin{figure*}
    \centering
    \includegraphics{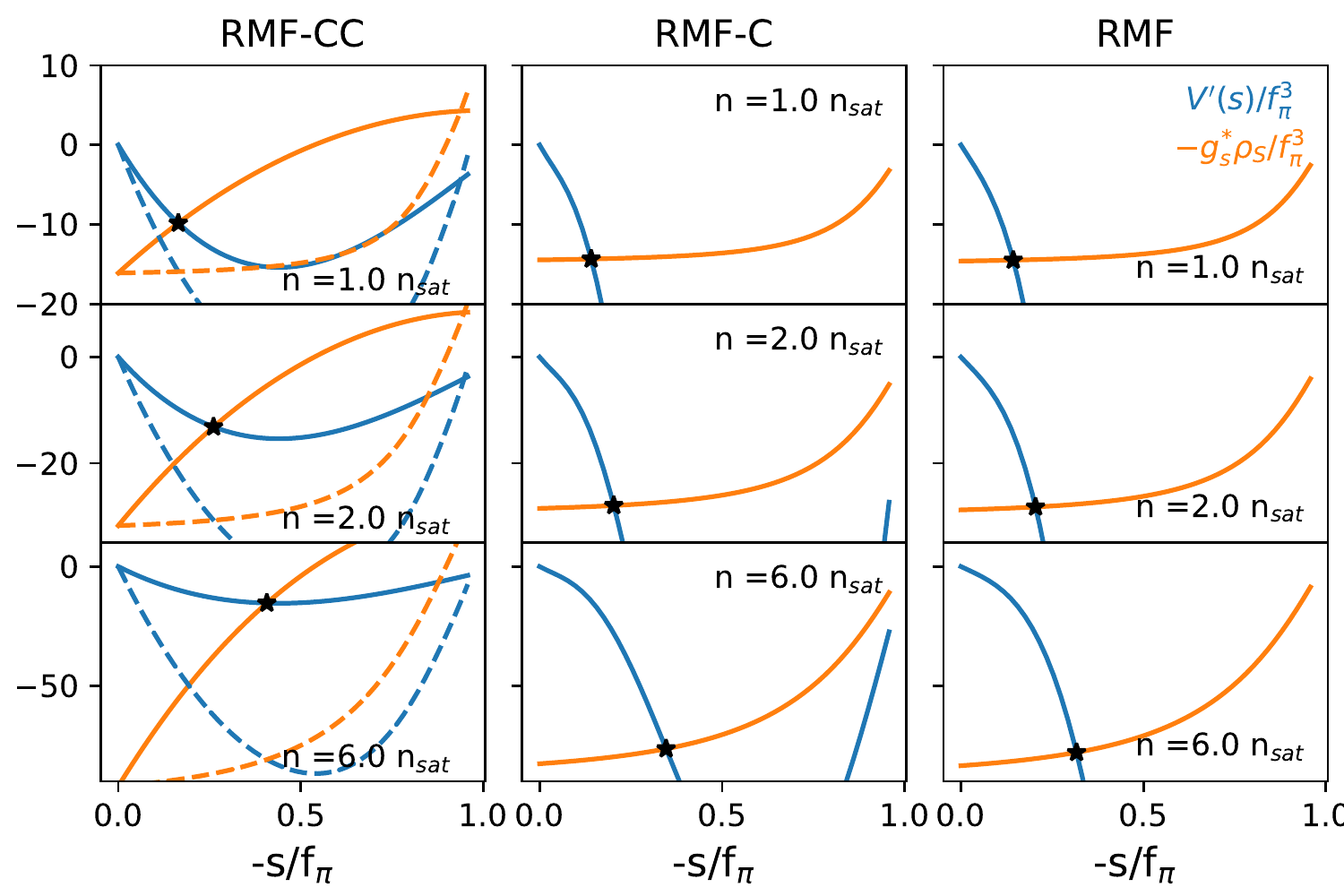}
    \caption{Analysis of the equation of motion of the scalar field. The different rows correspond to different densities. Dashed lines correspond to the equation of motion written for RMF-CC with the effective potential, see Eq.~\eqref{eq:sclar_eom_mod}.}
    \label{fig:Potential}
\end{figure*}

We now analyze in details the Equation of Motion (EoM) for the scalar field where the scalar potential plays naturally a crucial role. 
We will show that the anomaly of the scalar potential observed in the previous subsection for RMF-C and RMF models impacts the solution of the EoM. 

The EoM for the scalar field $s$ is, see~\ref{sec:a1}, 
\begin{equation}
    V'(s)=-g^*_S \rho_S\qquad\hbox{with}\qquad g^*_S=\frac{\partial M_N(s)}{\partial s},
    \label{eq:scalar_eom}
\end{equation}
where $\rho_S$ is the scalar density. Note that for RMF-C and RMF models, $g^*_s = g_s$ since $M_N(s)$ is simply linear in the field $s$.

In Fig. \ref{fig:Potential}, we represent the graphical solution of the EoM by drawing the two sides of the equation: $V'(s)$ is plotted as the solid blue lines and $-g^*_S \rho_S$ as the solid orange lines for the three models (by columns) and at different densities (by rows). 
The parameters of the scalar potential are taken to be the mean values reported in Table.~\ref{tab:g}, and similarly we consider the centroid of the PDFs for $g_s$ which are $11.10$, $9.98$ and $10.08$ for RMF-CC, RMF-C and RMF respectively.
In case of several solutions, the physical one is the smallest one and it is identified as a black star. 
We see that as the density increases, the solution for the $s$-field (abscissa of the black star) of the scalar EoM increases. It is interesting to note that for all three models, at a given density, the value of this solution is quite similar. This is due to the fact that the quantity $g_s s$ defines the in-medium Dirac mass (except for RMF-CC where the nucleon response is also included) which remains almost identical for the three models, see Fig.~\ref{fig:eff_mass} and $g_s$ does not differs by more than about 10\% between the different models, see Fig.~\ref{fig:scatter}. As a consequence the values of the field $s$ are very close between the three models considered here.

The absolute value of the y-coordinate of the solution ($\propto V^\prime(s)$) is however always smaller for RMF-CC compared to the RMF-C and RMF models. 
Since $V(s)$ is the vacuum chiral potential in the case of RMF-CC, the vertical position of the intersection point informs us about the derivatives of this potential for various values of the field $s$. In other words, for RMF-CC density scans the chiral potential function of $s$ and the in-medium effects are entirely captured by the nucleon response given by $g_s^*$.

The situation is however different for RMF-C and RMF models. These models share two important features: i) they do not incorporate explicitly the nucleon response as in RMF-CC, and ii) the scalar potential is determined from the fit to saturation properties. If the fit imposes a modification of the scalar potential making it different from the vacuum one, it is interpreted as an in-medium correction to the scalar potential. It is interesting to remark that the result of the fit, which is made differently for RMF-C and RMF, is to impose larger absolute values for $V^\prime(s)$ as function of $s$ compared to the vacuum values represented by RMF-CC model. As a consequence, the intersection points in RMF-C and RMF happen at larger absolute values compared to RMF-CC.
The vertical intersection point therefore informs us either on the role of the nucleon response in the EoM \eqref{eq:scalar_eom} (for RMF-CC), or on the in-medium modification of the scalar potential (for RMF-C and RMF).

At first sight, the models fitted to saturation and disregarding nucleon response (RMF-C and RMF) suggest large in-medium modification of the scalar potential, while the models considering the vacuum chiral potential complemented with nucleon response (RMF-CC) do not require any in-medium modification of the chiral potential.
One may however wonder to which extend these two opposite conclusions do not reflect a similar reality suggesting that the nucleon polarization may modify in an effective way the vacuum scalar potential. It may even be the dominant in-medium correction to the scalar EoM.

\begin{figure*}
    \centering
    \includegraphics[width=0.32\textwidth]{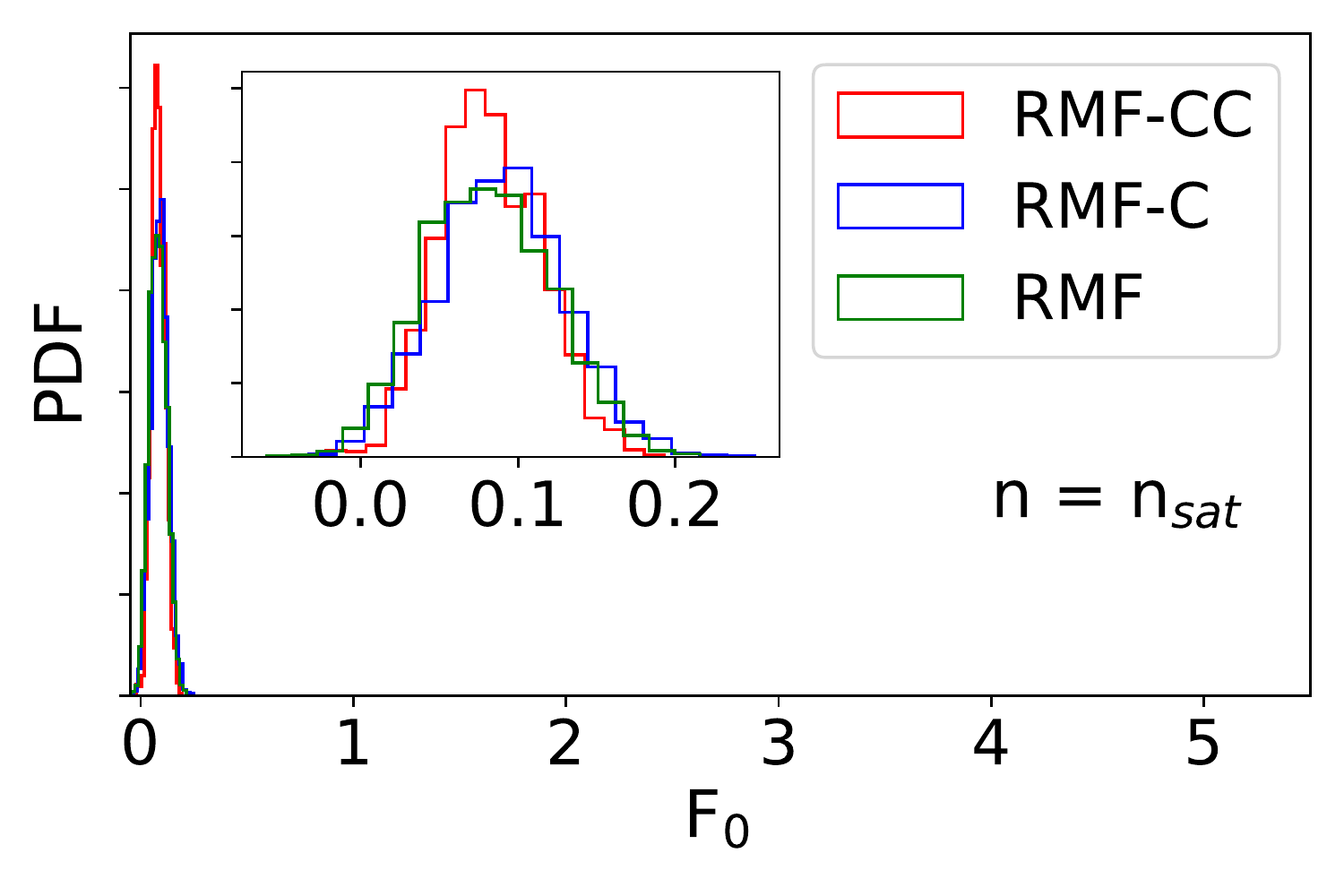}
    \includegraphics[width=0.32\textwidth]{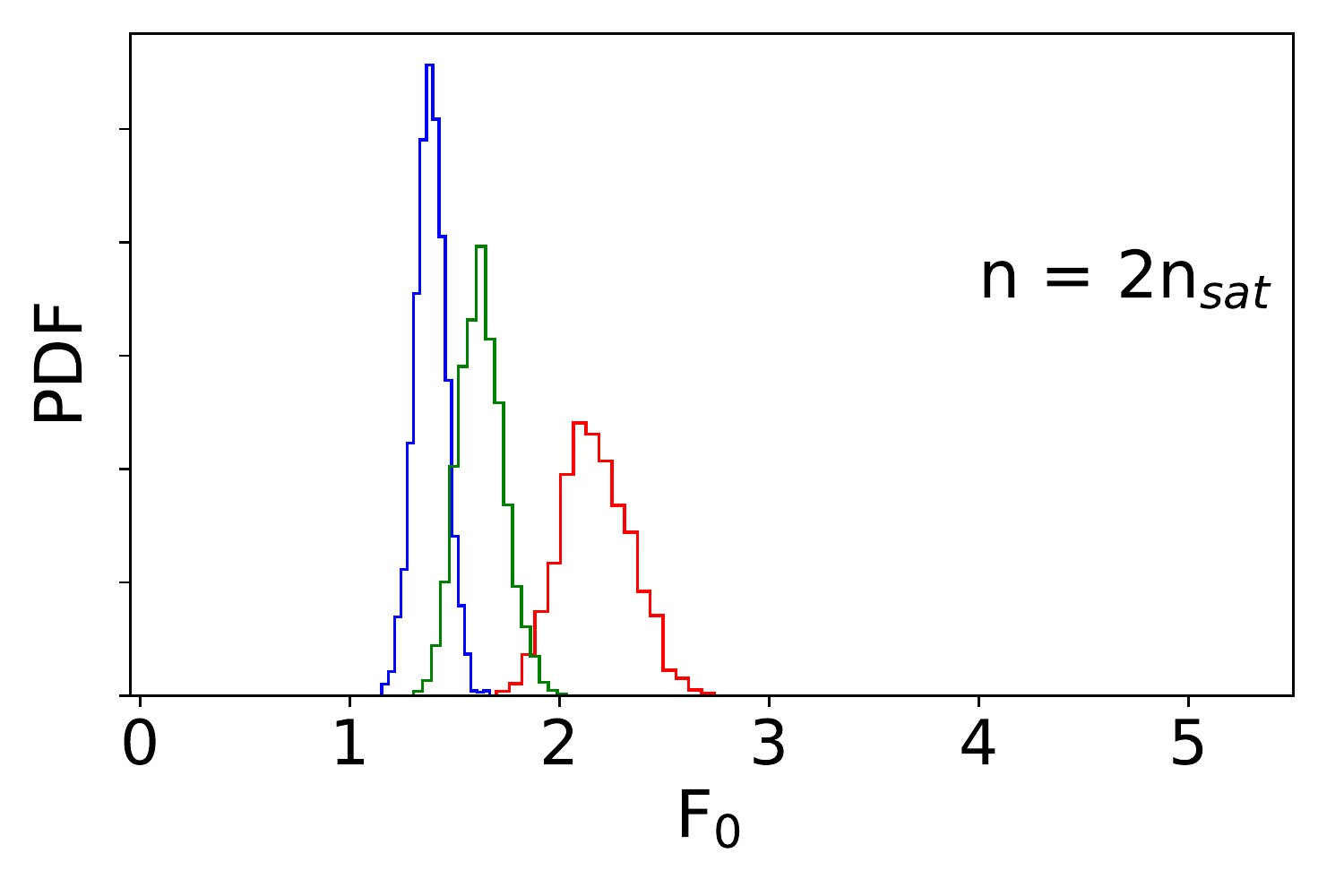}
    \includegraphics[width=0.32\textwidth]{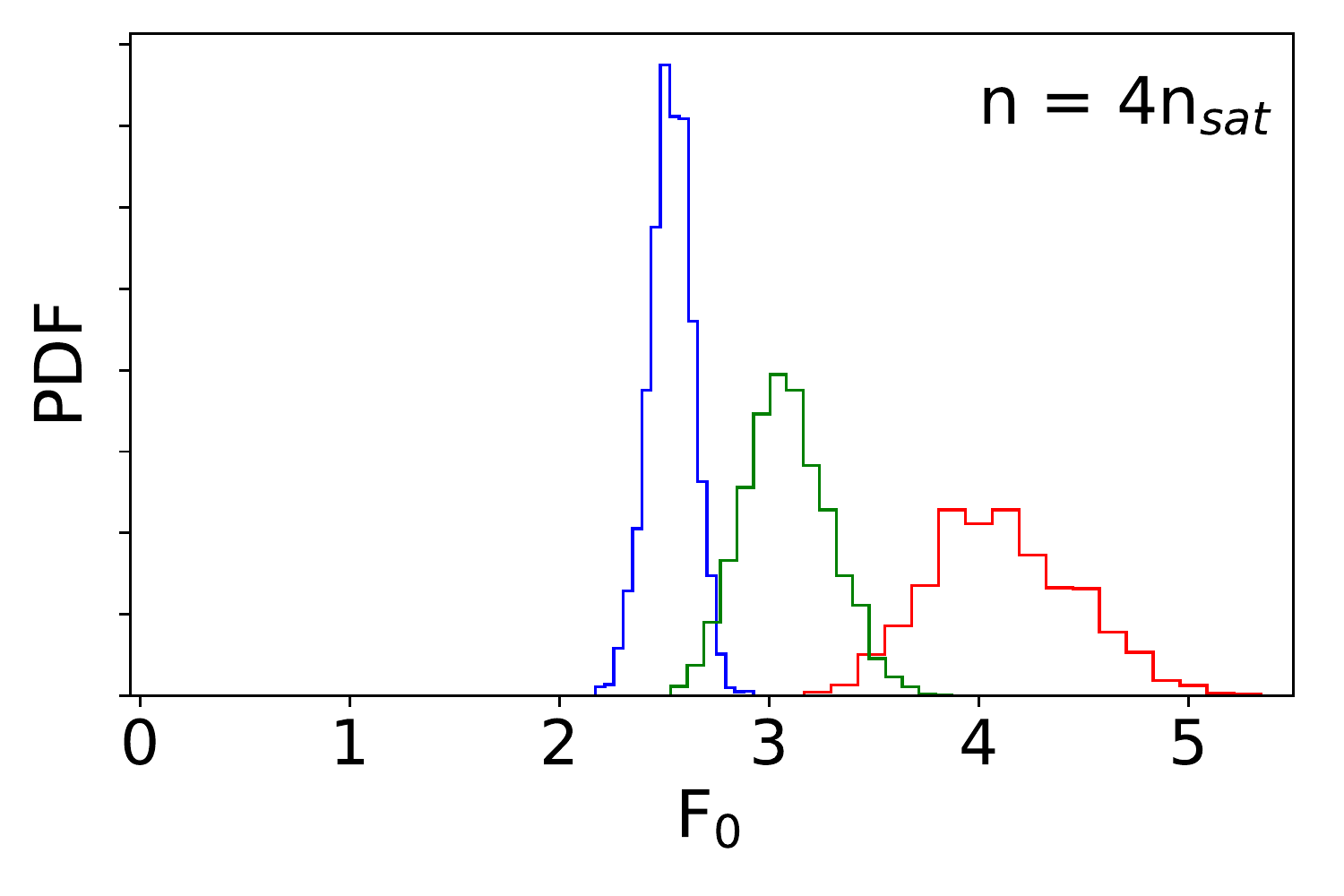}
    \caption{The Landau parameter evaluated at different densities. In the top panel, the inset shows a zoom of the region where the PDFs are nonzero.}
    \label{fig:Landau_parameter}
\end{figure*}

In order to address this question, we rewrite the EoM for RMF-CC to absorb the effects of the nucleon response in an effective scalar potential $\tilde{V}'(s)=V'(s) + g^*_S \rho_S - g_S \tilde{\rho}_S$, as 
\begin{equation}
\label{eq:sclar_eom_mod}
\tilde{V}'(s) = - g_S \tilde{\rho}_S \, .
\end{equation}
This leaves the standard scalar coupling $g_S$ on the RHS of Eq.~\eqref{eq:sclar_eom_mod} (as in the other models). Note that this re-arrangement has been done by ensuring that it is still the same self-consistent equation of motion that is being solved for the RMF-CC. Finally, $\tilde{\rho}_S$ denotes that the dependence of the scalar density on $s$ via the nucleon mass $M_N(s)$ is obtained by using a linear relation for $M_N(s)$ (as in RMF-C and RMF) and not the non-linear one used in RMF-CC.
In this way, Eq.~\eqref{eq:sclar_eom_mod} is formally equivalent to the EoM \eqref{eq:scalar_eom} solved for RMF-C and RMF models.

In Fig.~\ref{fig:Potential}, the left column (for RMF-CC) displays dashed blue and dashed orange lines corresponding to the graphical solution of Eq.~\eqref{eq:sclar_eom_mod} in terms of the effective potential $\tilde{V}'(s)$.
With such a construction, we see that the dashed blue line intersects the dashed orange line for lager absolute values of $\tilde{V}'(s)$, similar to RMF-C and RMF. This clearly demonstrates that the smaller absolute values of $V'(s)$ obtained for RMF-CC is a consequence of the inclusion of the scalar response of the nucleon. 
In other words, the nucleon polarisation captures most of the in-medium correction to the vacuum scalar potential.

While Fig.~\ref{fig:Potential} clearly demonstrates that the nucleon polarisation is the dominant in-medium correction to the scalar EoM, a similar conclusion may have been obtained in the previous section based on the values for $c_2$ given in Tab.~\ref{fig:scatter}.
In the RMF-CC model, the (positive) $c_2$ parameter controls the magnitude of the above mentioned attractive tadpole diagram which destroys saturation. For hierarchically ordered scalar potentials, it has been shown that after an appropriate shift of the scalar field, $\sigma_W=s+(\kappa_{NS}/2 g_s)s^2$, where the term $\propto \kappa_{NS}$ represents the nucleon response, the Dirac mass of the nucleon become $M_N(s)=M_N+g_s \sigma_W$ and the nucleon polarizability renormalizes the cubic term of the scalar potential as $c_2\, (1-2 C)\simeq -2\, c_2$ if $C\simeq 1.5$~\cite{Ericson2007} . One sees that this is qualitatively compatible with the values of $c_2$ for RMF-CC and NL3 quoted in Tab.~\ref{tab:g}. One can thus remark that the negative value of $c_2$ in the original NL3 model Ref.~\cite{Lalazissis1996} simulates in an effective way the nucleon response. Note also that this discussion is not applicable to the RMF model since its scalar potential displays an anomalous behaviour. However it is interesting to note that while having a positive centroid for $c_2$, negative values for $c_2$ are also allowed in the PDF for the RMF model.

In conclusion, we have shown that the in-medium modification of the scalar potential which is captured in RMF-C and RMF models by the fit to saturation properties can also be simulated in RMF-CC by a single in-medium term in the Lagrangian: the nucleon response generated by the coupling of the constituent quarks to the large scalar field at finite density. The nucleon response effect, being characterized by a single coupling constant ($\kappa_{NS}$ or $C$) in the RMF-CC Lagrangian represents therefore a very economical way to capture in-medium correction to the scalar EoM, on top of being well motivated from a microscopic viewpoint. In RMF-CC the chiral potential at finite density is identical to the vacuum one and one could interpret the solution of the scalar EoM as a scan of $V^\prime(s)$ at different values of $s$. This latter point suggests that the solution of the EoM at finite density may be a way to probe the properties of the chiral potential in vacuum.

\section{Excitations in dense matter}
\label{sec:Landau}

In the previous section, we have illustrated the equivalence between actual in-medium modification of the scalar potential guided by the fit to saturation properties (as in RMF-C and RMF) and in-medium effect of the nucleon polarization (as in RMF-CC). We have also suggested that the nucleon polarization is an economical way to treat in-medium correction to the scalar EoM. One may however wonder if the effect of the nucleon polarization could influence other properties in medium. It is therefore natural to come to the exploration of the excitation spectrum of dense matter.

We limit ourself to the scalar-isoscalar excitation channel, which is determined by the scalar-isoscalar Landau parameter $F_0$ at low excitation energy (and zero momentum transferred).
Following Ref.~\cite{Chanfray2005}, we have computed the fully relativistic Landau parameter $F_0$ for the three models (see~\ref{sec:a1} for the derivation). The final expression is 
\begin{align*}
 F_0 = & N_{0R}\Bigg(\frac{g^2_\omega}{m^2_\omega}  
 -\,\frac{g^{*2}_S}{m^{*2}_\sigma}\,\left(\frac{M_N(s)}{E_F}\right)^2 \\
 &\hspace{2cm}\times 
 \left[1\,+\,\frac{g^{*2}_S}{m^{*2}_\sigma}\,I_3(k_F)\right]^{-1}\Bigg),  
\end{align*}
where the meaning of the various symbols are explained in~\ref{sec:a1}. 
The nucleon polarization appears in $F_0$ through the effective scalar coupling constant $g_S^*$.

Fig.~\ref{fig:Landau_parameter} shows the Landau parameter $F_0$, for the three models evaluated at three different densities, $n_{\sat}$, $2n_{\sat}$ and $4n_{\sat}$. We see that at saturation density, the results agree for the three models, as it could be expected because the three models are calibrated such that they reproduce the same incompressibility modulus ($K_\sat$), see Eq.~\eqref{eq:app:ksat}.
However, at larger densities, the RMF-CC models predict larger values of $F_0$, followed by RMF and then RMF-C. At $4n_\sat$ the values for $F_0$ suggested by RMF-CC are almost twice the ones predicted by RMF-C.

This distinction in the predictions of $F_0$ by the three models at large densities may have important phenomenological consequences for dense matter in neutron stars. As an example, since it modifies the nuclear response functions, it may have implications for neutrino scattering and other phenomena in the core of neutron stars.
Additionally, it may also modify the properties of the dense fire-ball produced by relativistic heavy-ion collision.

\section{The symmetry Energy}
\label{sec:symmetry_energy}

\begin{figure}
    \centering
    \includegraphics[scale=0.5]{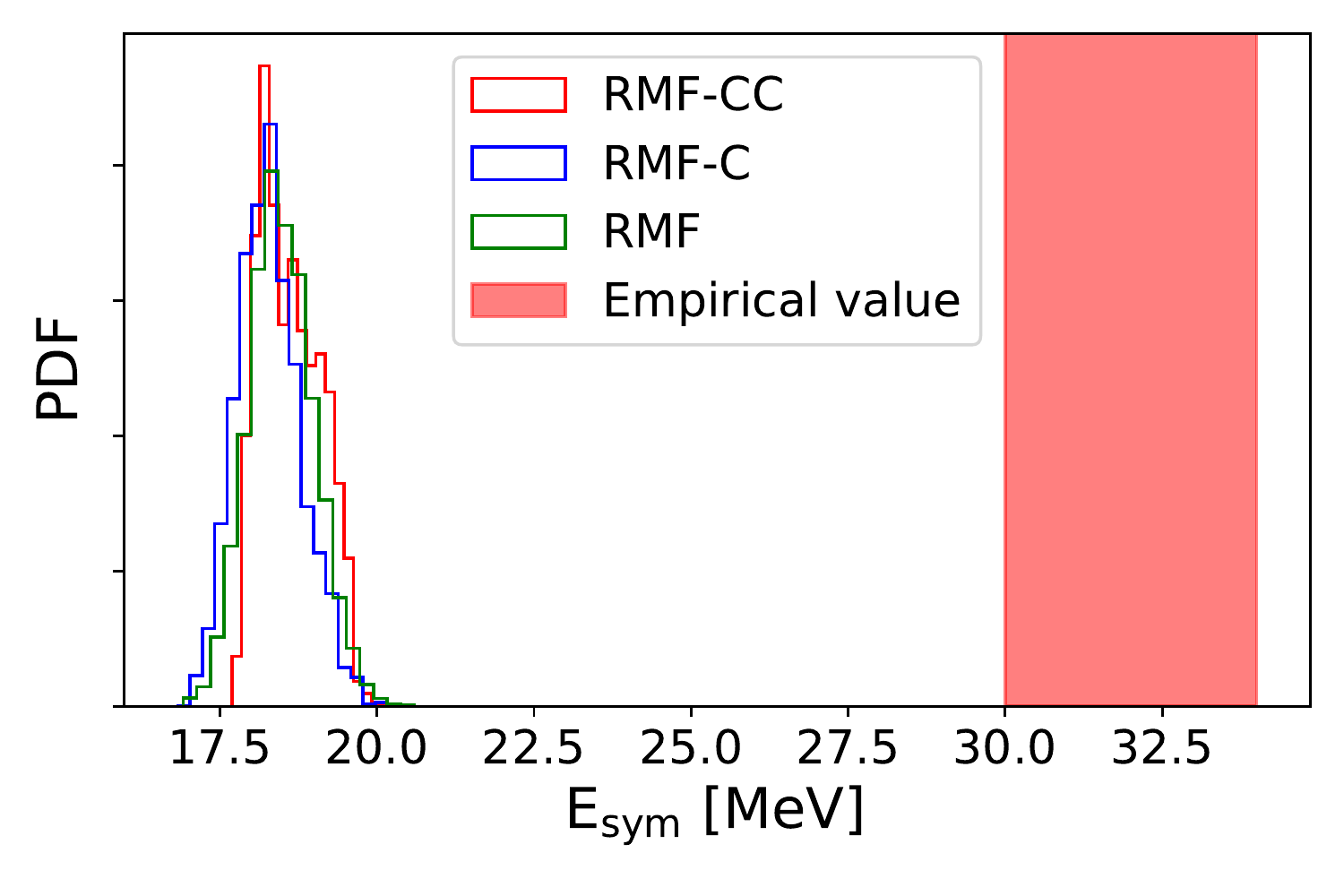}
    \includegraphics[scale=0.5]{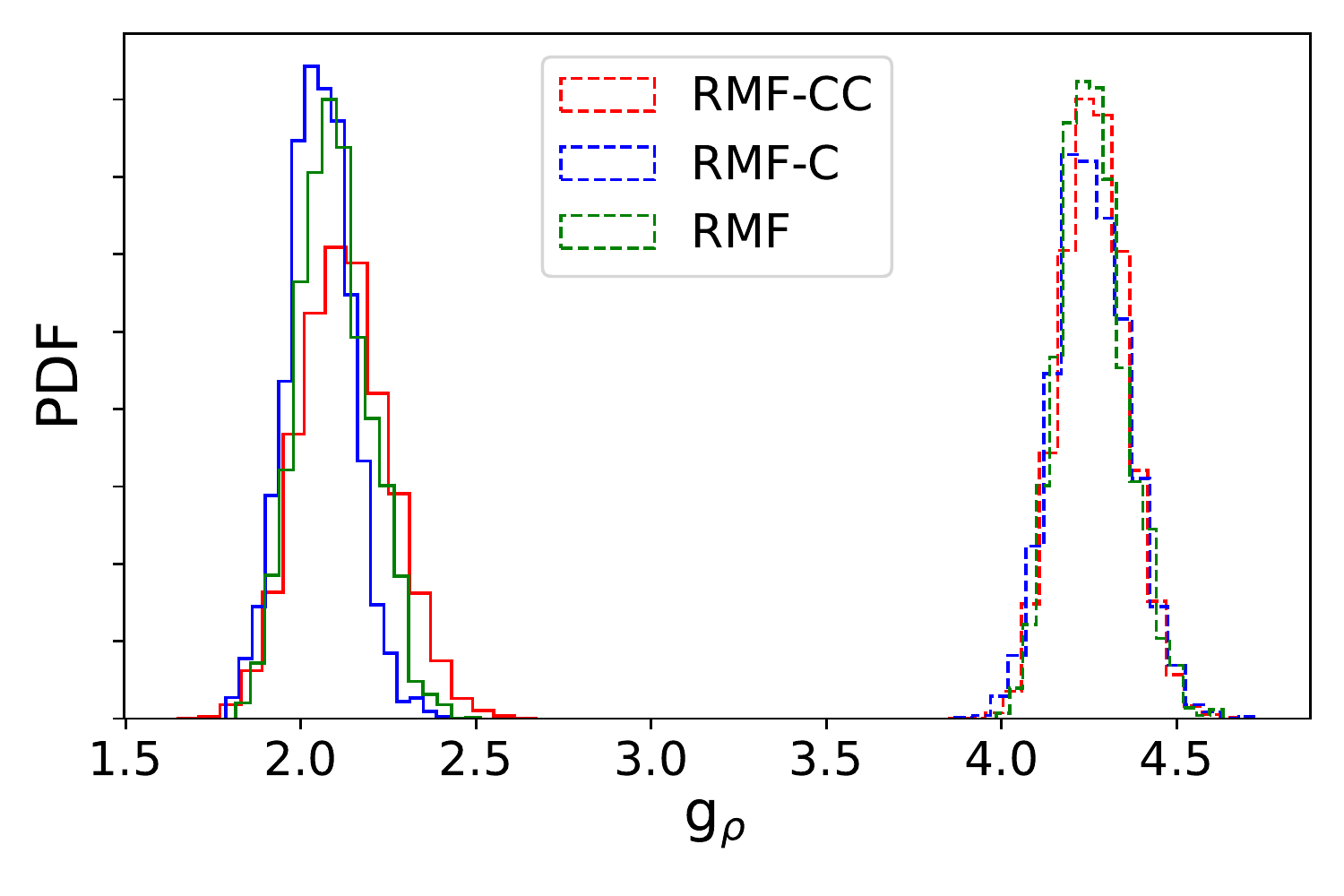}
    \caption{(TOP) The Symmetry Energy predicted by the three models assuming the quark model relation between $g_\rho$ and $g_\omega$. The empirical value (that can be obtained by breaking the quark model relation) is shaded in red. (BOTTOM) The coupling constant predicted by the quark model is shown as solid lines and the coupling required to obtain the empirical value is shown as dashed lines.}
    \label{fig:E_sym}
\end{figure}

In this paper, we have restricted our many-body treatment to the Hartree approximation (classical fields) and to SM. In the future, we will also include the contribution of the Fock terms and we will explore asymmetric matter. At the Hartree level, the symmetry energy is however only determined by the $\rho$ vector iso-vector meson since the small contribution of the $\delta$ meson is neglected here. One obtains
\begin{equation}
E_\sym = \frac{k_F^2}{6\sqrt{k_F^2+M_N^2(s)}} + \frac{g_\rho^2}{2m_\rho^2} \rho \,
\end{equation}
where $g_\rho$ and $m_\rho$ are the coupling constant and the mass of the $\rho$ meson. If the quark model is assumed, then $g_\rho = g_\omega/3$. In this case, the predictions for the symmetry energy at saturation density $E_\sym$ is shown for the three models in the upper panel of Fig.~\ref{fig:E_sym}. The empirical value is shown as a red band. We see that there is a difference of about $12-15$ MeV (about half the expected value for $E_\sym$) between the predicted value and the empirical one. In the lower panel, in solid lines the value of $g_\rho$ assuming the quark model (as in the upper panel) is confronted to the dashed lines showing
the value of $g_\rho$ required to reproduce the empirical value for $E_\sym$.
There is a factor 2 difference between the solid and the dashed curves.

We thus see that setting $g_\rho \approx 4.25$ would be a simple way of obtaining the empirical value of $E_\sym$. However, we interpret the discrepancy between the quark model prediction and the empirical value for $E_\sym$ as originating from the correlations beyond the Hartree approximation. 
In a future work, we will illustrate this point by adding to the present modeling the contribution of the Fock term, without modification of the fitting procedure. Preliminary results presented in Ref.~\cite{Massot2008} give us confidence in our interpretation.

\section{Conclusions}
\label{sec:conclusion}

In this paper we have compared several classes of relativistic models applied to the description of nuclear matter. A systematical analysis was performed by ensuring a democratic treatment of the three models: we constrained the three models to agree with each other in the vicinity of saturation density of SM. In particular, the fit of the RMF-CC parameters to SM properties was performed by incorporating the parameter uncertainty, in the L-QCD parameters for instance, and propagating them in the predictions in dense matter using Bayesian statistics.
In doing so, we directly connect RMF-CC with the underlying QCD theory and explore how uncertainties in this link propagate as function of the density. The fit of RMF-C and RMF models to SM properties predicted by RMF-CC was performed in a consistent Bayesian manner which allows us to properly explore the uncertainties in the empirical knowledge of nuclear matter saturation.

Examining various aspects of the models and their predictions at different densities we have shown that the scalar nucleon response is a microscopically justified and an economical way to incorporate in-medium corrections to the scalar EoM. 
In RMF-CC the modification of the effective potential at high density is driven by a microscopic mechanism, while in RMC-C and RMF approaches the scalar potential already encompasses finite density properties at saturation, that are simply extrapolated to high densities. 
In addition, we have shown that if the nucleon response is neglected, the scalar potential become anomalous since the hierarchy in the orders of $s$ is not respected as it is expected in a many-body framework. Moreover the ground state and excited states in the scalar-isoscalar channel are predicted to be noticeably different among the various classes of relativistic approaches as the density increases (2 to $4 n_\sat$).

Finally, phase transitions are expected to occur in the very dense matter found in the core of massive neutron stars. These phase transitions could lead to the appearance of other (non-nucleonic) hadronic degrees of freedom such as pion and kaon condensates as well as hyperons. Other phenomena such as chiral symmetry restoration and transition to deconfined quark matter might also take place. The exploration of these two possibilities will also be very interesting to investigate in the future using the models developed in this work, since they incorporate chiral symmetry and confinement.

\begin{acknowledgements}
G.C., H.H., J.M. and R.S. are supported by the CNRS IN2P3 NewMAC project,
and benefit from PHAROS COST Action MP16214. 
This work is supported by the STRONG-2020 network from the European
Union's Horizon 2020 research and innovation program under grant agreement No. 824093 (H.H.);
the LABEX Lyon Institute of Origins (ANR-10-LABX-0066) of the
\textsl{Universit\'e de Lyon} for its financial support within the
program \textsl{Investissements d'Avenir} (ANR-11-IDEX-0007) of the
French government operated by the National Research Agency (ANR).
\end{acknowledgements}

\appendix 
	
\section{Compressibility modulus and Landau parameter $F_0$ in relativistic theory}
\label{sec:a1}

The nuclear matter energy density in the RMF theories discussed in this paper can be written as :
\begin{eqnarray}
 \varepsilon &=&\int\frac{4 d^3 k}{(2\pi)^3}\left( \sqrt{k^2+M_N^2(s)} \,+\,g_\omega\,\omega_0\right) \,\Theta(k_F-k)\nonumber\\
 && \hspace{2cm}+\,V(s)\,-\,\frac{1}{2}\,m^2_\omega\,\omega_0^2, 
\end{eqnarray}
where the scalar and vector field are obtained from the equations of motion : 
\begin{eqnarray}
&&m^2_\omega\,\omega_0=g_\omega\,\rho\\
&&V'(s)=-g^*_S \rho_S\qquad\hbox{with}\qquad g^*_S=\frac{\partial M_N(s)}{\partial s}.\label{EOMS}
\end{eqnarray}
In the following we will make use of the following relations or definitions: 
\begin{equation}
\rho=\frac{1}{3} \frac{2\, k^3_F}{\pi^2}, \quad \frac{\partial k_F}{\partial\rho}=\frac{1}{3}\frac{k_F}{\rho}\quad
N_{0R}=\frac{2\, k_F E_F}{\pi^2},
\end{equation}
where $E_F=\sqrt{k^2_F+M_N^2(s)}$ is the Fermi energy and $N_{0R}$ is the density of states on the (relativistic) Fermi surface. The first derivative with respect to the density of the energy density is obtained using equations of motion with the result:
\begin{equation}
\frac{\partial\varepsilon}{\partial\rho} =E_F\,+\,\frac{g^2_\omega}{ m^2_\omega}\,\rho\equiv\mu .   
\end{equation}
The second derivative is:
\begin{equation}
\frac{\partial^2\varepsilon}{\partial\rho^2} =\left[\frac{k_F}{E_F}\,+\,\frac{M_N(s)}{E_F}\,g^*_S\,\frac{\partial s}{\partial k_F}\right]\frac{\partial k_F}{\partial\rho}\,+\,\frac{g^2_\omega}{m^2_\omega}.
\end{equation}
The derivative of $s$ with respect to the Fermi momentum is obtained by taking the derivative of the equation of motion (Eq. \ref{EOMS}):
\begin{equation}
V''(s)\frac{\partial s}{\partial k_F}=-\tilde\kappa_{NS}\,\frac{\partial s}{\partial k_F}\rho_S\,+\,g ^*_S\rho_S\,\quad\hbox{with}\quad \tilde\kappa_{NS}=\frac{\partial g ^*_S}{\partial s}.
\end{equation}
We introduce an effective sigma meson mass, such as  $m^{*2}_\sigma=m^2_\sigma +\tilde\kappa_{NS}\,\rho_S $, to obtain:
\begin{equation}
\frac{\partial s}{\partial k_F}=-\frac{1}{m^{*2}_\sigma}\,\frac{\partial \rho_S}{\partial k_F}\label{DEV1}
\end{equation}
and the derivative of the scalar density has the form:
\begin{eqnarray}
\frac{\partial \rho_S}{\partial k_F} &=&\frac{\partial}{\partial k_F}\left[\int\frac{4 d^3 k}{(2\pi)^3}\frac{M_N(s)}{\sqrt{k^2+M_N^2(s)}} \,\Theta(k_F-k) \right]  \nonumber\\ 
&=&\frac{\partial \rho}{\partial k_F}\frac{M_N(s)}{E_F}\,+\,I_3(k_F)\,g^*_S\,\frac{\partial s}{\partial k_F}\nonumber\\
\hbox{with}&& I_3=\int\frac{4 d^3 k}{(2\pi)^3}\frac{k^2}{\left(k^2+M_N^2(s)\right)^{3/2}}\Theta(k_F-k). \label{DEV2}
\end{eqnarray}
Combining Eqs. (\ref{DEV1}) and (\ref{DEV2}), we obtain:
\begin{eqnarray}
\frac{\partial \rho_S}{\partial \rho}&=&\frac{\partial \rho_S}{\partial k_F}\,\frac{\partial k_F}{\partial \rho}\nonumber\\
&=&-\frac{g^{*2}_S}{m^{*2}_\sigma}\,\frac{M_N(s)}{E_F}\left[1\,+\,\frac{g^{*2}_S}{m^{*2}_\sigma}\,I_3(k_F)\right]^{-1}.
\end{eqnarray}
Using the previous results, the compressibility modulus can be written in the following form:
\begin{equation}
K_\sat = 9\rho\,\frac{\partial^2\varepsilon}{\partial\rho^2} =\frac{3\,k^2_F}{E_F}\left(1\,+\,F_0\right), 
\label{eq:app:ksat}
\end{equation}
which depends on the relativistic generalization of the Landau parameter $F_0$:
\begin{multline}
 F_0 = N_{0R}  \ \Biggl(\frac{g^2_\omega}{m^2_\omega}\,-\,  \frac{g^{*2}_S}{m^{*2}_\sigma}\,\ \Biggl(\frac{M_N(s)}{E_F}\ \Biggr)^2   \\
 \ \Biggl[1\,+\,\frac{g^{*2}_S}{m^{*2}_\sigma}\,I_3(k_F)\ \Biggr]^{-1}\ \Biggr).  
\end{multline}
This result derived in a different manner has been quoted in \cite{Chanfray2005} but omitting the (small) correction arising from the $I_3$ integral. Notice that, as demonstrated in Ref. \cite{Chanfray2003},  $g^{*2}_S\, I_3(k_F)$ corresponds to the nuclear response associated with $N\bar  N$ excitation. Also notice that our result coincides with the one derived by T. Matsui \cite{Matsui}  but in the absence of medium modification (i.e., in the absence of the nucleon susceptibility term) of the scalar mass and coupling constant.

\bibliographystyle{spphys}       
\bibliography{biblio}

\begin{thebibliography}{10}
\providecommand{\url}[1]{{#1}}
\providecommand{\urlprefix}{URL }
\expandafter\ifx\csname urlstyle\endcsname\relax
  \providecommand{\doi}[1]{DOI \discretionary{}{}{}#1}\else
  \providecommand{\doi}{DOI \discretionary{}{}{}\begingroup
  \urlstyle{rm}\Url}\fi

\bibitem{CBM2017}
T.~Ablyazimov, et~al., Eur. Phys. J. A \textbf{53}(3), 60 (2017).
\newblock \doi{10.1140/epja/i2017-12248-y}

\bibitem{Weinberg}
S.~Weinberg, Phys. Lett. B \textbf{251}, 288 (1990).
\newblock \doi{10.1016/0370-2693(90)90938-3}

\bibitem{Tews:2018kmu}
I.~Tews, J.~Carlson, S.~Gandolfi, S.~Reddy, Astrophys. J. \textbf{860}(2), 149
  (2018).
\newblock \doi{10.3847/1538-4357/aac267}

\bibitem{Hebeler2013}
K.~Hebeler, J.M. Lattimer, C.J. Pethick, A.~Schwenk, Astrophys. J.
  \textbf{773}, 11 (2013).
\newblock \doi{10.1088/0004-637X/773/1/11}

\bibitem{Chanfray2001}
G.~Chanfray, M.~Ericson, P.A.M. Guichon, Phys. Rev. C \textbf{63}, 055202
  (2001).
\newblock \doi{10.1103/PhysRevC.63.055202}

\bibitem{Chanfray2005}
G.~Chanfray, M.~Ericson, Eur. Phys. J. A \textbf{25}, 151 (2005).
\newblock \doi{10.1140/epja/i2005-10074-6}

\bibitem{SerotWalecka1986}
B.D. Serot, J.D. Walecka, Adv. Nucl. Phys. \textbf{16}, 1 (1986)

\bibitem{Walecka1997}
B.D. Serot, J.D. Walecka, Int. J. Mod. Phys. E \textbf{6}, 515 (1997).
\newblock \doi{10.1142/S0218301397000299}

\bibitem{Wetterich}
S.~Floerchinger, C.~Wetterich, Nucl. Phys. A \textbf{890-891}, 11 (2012).
\newblock \doi{10.1016/j.nuclphysa.2012.07.009}

\bibitem{Drews2013}
M.~Drews, T.~Hell, B.~Klein, W.~Weise, Phys. Rev. D \textbf{88}(9), 096011
  (2013).
\newblock \doi{10.1103/PhysRevD.88.096011}

\bibitem{Fraga2018}
E.S. Fraga, M.~Hippert, A.~Schmitt, Phys. Rev. D \textbf{99}(1), 014046 (2019).
\newblock \doi{10.1103/PhysRevD.99.014046}

\bibitem{Schmitt2020}
A.~Schmitt, Phys. Rev. D \textbf{101}(7), 074007 (2020).
\newblock \doi{10.1103/PhysRevD.101.074007}

\bibitem{Demorest:2010}
P.~Demorest, T.~Pennucci, S.~Ransom, M.~Roberts, J.~Hessels, Nature
  \textbf{467}, 1081 (2010).
\newblock \doi{10.1038/nature09466}

\bibitem{Antoniadis:2013pzd}
J.~Antoniadis, P.C. Freire, N.~Wex, T.M. Tauris, R.S. Lynch, et~al., Science
  \textbf{340}, 6131 (2013).
\newblock \doi{10.1126/science.1233232}

\bibitem{Cromartie:2019}
H.T. Cromartie, et~al., Nature Astron. \textbf{4}(1), 72 (2019).
\newblock \doi{10.1038/s41550-019-0880-2}

\bibitem{Fonseca:2021wxt}
E.~Fonseca, et~al., Astrophys. J. Lett. \textbf{915}(1), L12 (2021).
\newblock \doi{10.3847/2041-8213/ac03b8}

\bibitem{Miller:2021qha}
M.C. Miller, et~al., Astrophys. J. Lett. \textbf{918}(2), L28 (2021).
\newblock \doi{10.3847/2041-8213/ac089b}

\bibitem{Riley:2021pdl}
T.E. Riley, et~al., Astrophys. J. Lett. \textbf{918}(2), L27 (2021).
\newblock \doi{10.3847/2041-8213/ac0a81}

\bibitem{Bedaque:2014sqa}
P.~Bedaque, A.W. Steiner, Phys. Rev. Lett. \textbf{114}(3), 031103 (2015).
\newblock \doi{10.1103/PhysRevLett.114.031103}

\bibitem{Somasundaram:2021clp}
R.~Somasundaram, I.~Tews, J.~Margueron, To be submitted. arXiv:2112.08157
  [nucl-th]  (2021)

\bibitem{Wong:1998ex}
S.S.M. Wong, \emph{{Introductory nuclear physics}} (Wiley, 1998)

\bibitem{Yndurain2006}
F.J. Yndurain, \emph{{The Theory of Quark and Gluon Interactions}}.
\newblock Theoretical and Mathematical Physics (Springer, Berlin, Germany,
  2006).
\newblock \doi{10.1007/3-540-33210-3}

\bibitem{Guichon1988}
P.A.M. Guichon, Phys. Lett. B \textbf{200}, 235 (1988).
\newblock \doi{10.1016/0370-2693(88)90762-9}

\bibitem{Yukawa1935}
H.~Yukawa, Proc. Phys. Math. Soc. Jap. \textbf{17}, 48 (1935).
\newblock \doi{10.1143/PTPS.1.1}

\bibitem{Erkelenz1971}
K.~Erkelenz, Phys. Rept. \textbf{13}, 191 (1974).
\newblock \doi{10.1016/0370-1573(74)90008-8}

\bibitem{Lalazissis1996}
G.A. Lalazissis, J.~Konig, P.~Ring, Phys. Rev. C \textbf{55}, 540 (1997).
\newblock \doi{10.1103/PhysRevC.55.540}

\bibitem{Brockmann1984}
R.~Brockmann, R.~Machleidt, Phys. Lett. B \textbf{149}, 283 (1984).
\newblock \doi{10.1016/0370-2693(84)90407-6}

\bibitem{TerHaar1986}
B.~Ter~Haar, R.~Malfliet, Phys. Rept. \textbf{149}, 207 (1987).
\newblock \doi{10.1016/0370-1573(87)90085-8}

\bibitem{Typel1999}
S.~Typel, H.H. Wolter, Nucl. Phys. A \textbf{656}, 331 (1999).
\newblock \doi{10.1016/S0375-9474(99)00310-3}

\bibitem{vanDalen2011}
E.N.E. van Dalen, H.~Muther, Phys. Rev. C \textbf{84}, 024320 (2011).
\newblock \doi{10.1103/PhysRevC.84.024320}

\bibitem{Lalazissis2005}
G.A. Lalazissis, T.~Niksic, D.~Vretenar, P.~Ring, Phys. Rev. C \textbf{71},
  024312 (2005).
\newblock \doi{10.1103/PhysRevC.71.024312}

\bibitem{Long2007}
W.H. Long, H.~Sagawa, N.~Van~Giai, J.~Meng, Phys. Rev. C \textbf{76}, 034314
  (2007).
\newblock \doi{10.1103/PhysRevC.76.034314}

\bibitem{Ripka1997}
G.~Ripka, \emph{{Quarks bound by chiral fields: The quark-structure of the
  vacuum and of light mesons and baryons}} (Clarendon Press, 1997)

\bibitem{Birse94}
M.C. Birse, Phys. Rev. C \textbf{53}, R2048 (1996).
\newblock \doi{10.1103/PhysRevC.53.R2048}

\bibitem{Chanfray2011}
G.~Chanfray, M.~Ericson, Phys. Rev. C \textbf{83}, 015204 (2011).
\newblock \doi{10.1103/PhysRevC.83.015204}

\bibitem{Chanfray2006}
G.~Chanfray, D.~Davesne, M.~Ericson, M.~Martini, Eur. Phys. J. A \textbf{27},
  191 (2006).
\newblock \doi{10.1140/epja/i2005-10245-5}

\bibitem{Massot2008}
E.~Massot, G.~Chanfray, Phys. Rev. C \textbf{78}, 015204 (2008).
\newblock \doi{10.1103/PhysRevC.78.015204}

\bibitem{Margueron2018}
J.~Margueron, R.~Hoffmann~Casali, F.~Gulminelli, Phys. Rev. C \textbf{97}(2),
  025805 (2018).
\newblock \doi{10.1103/PhysRevC.97.025805}

\bibitem{LWY2003}
D.B. Leinweber, A.W. Thomas, R.D. Young, Phys. Rev. Lett. \textbf{92}, 242002
  (2004).
\newblock \doi{10.1103/PhysRevLett.92.242002}

\bibitem{Boguta83}
J.~Boguta, Phys. Lett. B \textbf{120}, 34 (1983).
\newblock \doi{10.1016/0370-2693(83)90617-2}

\bibitem{KM74}
A.K. Kerman, L.D. Miller, in \emph{{1974 PEP Summer Study}} (1974)

\bibitem{BT01}
W.~Bentz, A.W. Thomas, Nucl. Phys. A \textbf{696}, 138 (2001).
\newblock \doi{10.1016/S0375-9474(01)01119-8}

\bibitem{C03}
G.~Chanfray, Nucl. Phys. A \textbf{721}, 76 (2003).
\newblock \doi{10.1016/S0375-9474(03)01019-4}

\bibitem{Guichon2004}
P.A.M. Guichon, A.W. Thomas, Phys. Rev. Lett. \textbf{93}, 132502 (2004).
\newblock \doi{10.1103/PhysRevLett.93.132502}

\bibitem{Chanfray2007}
G.~Chanfray, M.~Ericson, Phys. Rev. C \textbf{75}, 015206 (2007).
\newblock \doi{10.1103/PhysRevC.75.015206}

\bibitem{CP-PACS:2001vqx}
A.~Ali~Khan, et~al., Phys. Rev. D \textbf{65}, 054505 (2002).
\newblock \doi{10.1103/PhysRevD.65.054505}.
\newblock [Erratum: Phys.Rev.D 67, 059901 (2003)]

\bibitem{TGLY04}
A.W. Thomas, P.A.M. Guichon, D.B. Leinweber, R.D. Young, Prog. Theor. Phys.
  Suppl. \textbf{156}, 124 (2004).
\newblock \doi{10.1143/PTPS.156.124}

\bibitem{CSWSX95}
L.S. Celenza, C.M. Shakin, W.D. Sun, J.~Szweda, X.q. Zhu, Annals Phys.
  \textbf{241}, 1 (1995).
\newblock \doi{10.1006/aphy.1995.1053}

\bibitem{CWS01}
L.S. Celenza, H.~Wang, C.M. Shakin, Phys. Rev. C \textbf{63}, 025209 (2001).
\newblock \doi{10.1103/PhysRevC.63.025209}

\bibitem{Massot2009}
E.~Massot, G.~Chanfray, Phys. Rev. C \textbf{80}, 015202 (2009).
\newblock \doi{10.1103/PhysRevC.80.015202}

\bibitem{Massot2012}
E.~Massot, J.~Margueron, G.~Chanfray, EPL \textbf{97}(3), 39002 (2012).
\newblock \doi{10.1209/0295-5075/97/39002}

\bibitem{CM2020}
G.~Chanfray, J.~Margueron, Phys. Rev. C \textbf{102}(2), 024331 (2020).
\newblock \doi{10.1103/PhysRevC.102.024331}

\bibitem{Boguta89}
J.~Boguta, J.~Kunz, Nucl. Phys. A \textbf{501}, 637 (1989).
\newblock \doi{10.1016/0375-9474(89)90153-X}

\bibitem{Bhaduri1988}
R.K. Bhaduri, \emph{{Models of the nucleon: From quarks to soliton}}
  (Addison-Wesley, 1988)

\bibitem{Boguta1977}
J.~Boguta, A.R. Bodmer, Nucl. Phys. A \textbf{292}, 413 (1977).
\newblock \doi{10.1016/0375-9474(77)90626-1}

\bibitem{Glendenning1997}
N.K. Glendenning, \emph{{Compact stars: Nuclear physics, particle physics, and
  general relativity}} (Springer, 1997)

\bibitem{Ma2004}
Z.y. Ma, J.~Rong, B.Q. Chen, Z.Y. Zhu, H.Q. Song, Phys. Lett. B \textbf{604},
  170 (2004).
\newblock \doi{10.1016/j.physletb.2004.11.004}

\bibitem{Jaminon1989}
M.~Jaminon, C.~Mahaux, Phys. Rev. C \textbf{40}, 354 (1989).
\newblock \doi{10.1103/PhysRevC.40.354}

\bibitem{Ericson2007}
M.~Ericson, G.~Chanfray, Eur. Phys. J. A \textbf{34}, 215 (2007).
\newblock \doi{10.1140/epja/i2007-10498-x}

\bibitem{Chanfray2003}
G.~Chanfray, M.~Ericson, P.A.M. Guichon, Phys. Rev. C \textbf{68}, 035209
  (2003).
\newblock \doi{10.1103/PhysRevC.68.035209}

\bibitem{Matsui}
T.~Matsui, Nucl. Phys. A \textbf{370}, 365 (1981).
\newblock \doi{10.1016/0375-9474(81)90103-2}

\end{thebibliography}

\end{document}